\documentclass[a4paper,11pt]{article}
\usepackage{jcappub} 
\usepackage{lineno}

\title{\boldmath Radial dependence of galaxy and halo properties in cosmic voids: SDSS DR7 and the ELUCID constrained simulation}

\author{Youcai Zhang}

\affiliation{Shanghai Astronomical Observatory, \\
Nandan Road 80, Shanghai 200030, China}

\emailAdd{yczhang@shao.ac.cn}

\abstract{
Cosmic voids provide a setting for examining the interplay between
mass and environment in shaping galaxy and halo properties.
Using the SDSS DR7 volume-limited galaxy sample and the ELUCID
constrained simulation of the local large-scale structure, we investigate
the radial profiles of galaxy and subhalo properties across different
mass ranges, identifying voids independently in the observed galaxy
and simulated subhalo distributions. Both stellar mass and subhalo mass
increase from void interiors toward the void boundary, indicating that
lower-mass galaxies and subhaloes preferentially reside in void interiors.
The satellite fraction of subhaloes also increases systematically toward
the void boundary. Galaxy colour and specific star formation rate exhibit
radial variations that are weaker than their dependence on stellar mass,
while the spiral fraction shows little radial dependence at fixed stellar
mass. For dark-matter haloes, the apparent radial increase in size within
a broad high-mass bin largely disappears when narrower mass ranges are
considered. Halo concentration and spin show radial variations at fixed
mass, with higher concentrations and lower spins toward void interiors,
although the concentration trend is less pronounced in the extended
low-mass sample. The formation redshift of low-mass subhaloes also
increases toward the void boundary, whereas this dependence weakens
and eventually disappears at higher masses. Overall, the radial trends
in galaxy properties are generally weaker than their mass dependence,
while several subhalo properties retain measurable radial variations
at fixed mass.
}

\begin{document}
\maketitle
\flushbottom

\section{Introduction}
\label{sec_introduction}

Cosmic voids are among the largest structures in the cosmic web and
occupy a substantial fraction of the volume of the Universe
\cite{Cautun2014, Contarini2026, Song2026a, Song2026b}.
Their low matter density and distinct dynamical histories make them a
natural environment in which to investigate how galaxies and dark matter
structures form and evolve under conditions that differ from those in
denser regions
\cite{Sheth2004, Padilla2005, Hamaus2015, Hamaus2016, Argudo2026,
Moretti2026a, Moretti2026b, Moretti2026c, Rouse2026, Takadera2026}.
In particular, the spatial distributions and properties of galaxies and
dark-matter haloes within voids provide complementary views of the
structure and evolution of underdense regions
\cite{Colberg2005, Ricciardelli2014a, Nadathur2015a,Nadathur2015b,
Pollina2017, Pollina2019, Curtis2025}.

Observational studies have reported systematic variations in various
galaxy properties with environment, including stellar mass, colour,
star formation activity, and morphology
\cite{Beygu2016, Beygu2017, Habouzit2020, Florez2021}.
However, the environmental dependence of these properties can be
strongly influenced by the stellar-mass distribution of the galaxy
population. Since many galaxy properties depend strongly on stellar
mass, an apparent environmental trend in a full galaxy sample may
partly reflect changes in its stellar-mass composition rather than a
direct dependence on environment
\cite{Kuutma2017,Ricciardelli2017,Argudo2024}.
This raises the question of whether the observed variations in galaxy
properties reflect genuine environmental effects or are primarily
driven by the underlying mass dependence.

A complementary perspective is provided by cosmological
simulations, which allow the properties of the haloes and subhaloes
associated with galaxies to be studied directly. In addition to halo
mass, simulations provide access to quantities that are difficult to
measure observationally, such as halo formation time, internal size,
concentration, and angular momentum. These properties provide
information about the assembly history and internal structure of
dark-matter systems and offer a useful counterpart to the observed
properties of galaxies. Several of them are also known to depend on
the large-scale environment \cite{Hahn2007,Zhang2009, Zhang2021b}.

Previous studies have investigated environmental dependence by comparing
galaxies or haloes in voids with those in denser regions, or by
distinguishing different cosmic-web environments such as voids, walls,
filaments, and nodes \cite{Ganeshaiah2021,Hellwing2021,Hunde2025}. The
radial dependence of galaxy and halo properties within voids has also
been investigated, providing information on how these properties vary
from void interiors towards their boundaries
\cite{Habouzit2020, Rosas-Guevara2022, Rodriguez2023}. 
Extending this approach to multiple galaxy and subhalo properties, and examining 
their radial trends within restricted mass ranges, allows us to compare the 
environmental dependence of galaxies with that of their associated dark-matter structures.

An important challenge, however, is that conventional cosmological
simulations are not expected to reproduce the specific large-scale
structure observed in the nearby Universe. The ELUCID project addresses
this issue by reconstructing the initial density field of the local
Universe from observations and using the reconstructed field to
generate constrained $N$-body simulations
\cite{WangHuiyuan2009,WangHuiyuan2012,WangHuiyuan2013,
WangHuiyuan2014,WangHuiyuan2016}.
The resulting simulation reproduces key features of the observed 
large-scale structure within the reconstructed SDSS volume, while 
retaining the dynamical information provided by a cosmological $N$-body simulation.
This provides an observationally anchored framework in which the
properties of dark-matter haloes and subhaloes can be investigated
within a large-scale cosmic environment constrained by observations.
In this respect, combining the SDSS galaxy catalogue with the ELUCID
constrained simulation provides a complementary approach for studying
the galaxy--dark-matter connection in the local cosmic environment
\cite{Yang2018,Zhang2021a,Zhang2022,Zhang2024, Zhang2025,
Zhang2026a,Zhang2026b}.

In this work, we combine the Sloan Digital Sky Survey Data Release 7
(SDSS DR7) volume-limited galaxy sample with the ELUCID constrained
simulation to investigate the radial dependence of galaxy and
dark-matter subhalo properties within cosmic voids. Voids are identified
independently in the two datasets using the same \texttt{REVOLVER}
procedure \cite{Nadathur2019,Douglass2023}. For galaxies, we examine
stellar mass, colour, specific star formation rate (sSFR), morphology, and
size. For subhaloes, we investigate mass, formation redshift, size,
concentration, and spin.

We analyse the SDSS galaxy sample and two ELUCID subhalo samples: an
SDSS-matched sample with the same number density as the galaxy sample,
and an extended sample that includes lower-mass subhaloes. Each property
is measured for the full sample and within narrower stellar-mass or
subhalo-mass ranges, so that radial trends driven by mass can be
distinguished from those that remain at fixed mass. We characterize
these radial trends using the normalized void-centric radius
$r/R_{\rm v}$. For each radial bin, we first calculate the mean
property within each void and then average over voids, giving equal
weight to individual voids. The uncertainty is estimated from the
standard error of the mean of the void-level measurements.

Our main results show that the mean stellar mass of galaxies and the
mean mass of subhaloes both increase from void interiors towards void
boundaries. Galaxy colour and sSFR show radial
variations, but these trends are weaker than their dependence on
stellar mass, while the spiral fraction shows little radial dependence
at fixed stellar mass. For subhaloes, the apparent radial increase of
halo size in a broad high-mass bin disappears within narrower mass
ranges, whereas concentration, spin, satellite fraction, and the formation redshift of
low-mass subhaloes remain detectable within restricted mass ranges.
These results indicate which radial trends are closely linked to tracer 
mass and which remain after mass is restricted.

This paper is organized as follows. Section~\ref{sec_data} describes the SDSS
DR7 volume-limited galaxy sample, the ELUCID constrained simulation, the
tracer samples, the void catalogues, and the statistical measurements used
in this work. Section~\ref{sec_results} presents the results, including the radial
dependence of galaxy properties in Section~\ref{sec:galaxy_properties}
and subhalo properties in Section~\ref{sec:halo_properties}. The galaxy
analysis covers stellar mass, colour, sSFR,
morphology, and size, while the subhalo analysis considers subhalo mass,
satellite fraction, formation redshift, size, concentration, and spin.
Section~\ref{sec_conclusions} summarizes the main results and
conclusions.

Throughout this paper, we adopt the cosmological parameters used in the
ELUCID simulation: $\Omega_m = 0.258$, $\Omega_b = 0.044$,
$\Omega_\Lambda = 0.742$, $h = 0.72$, $n_s = 0.963$, and
$\sigma_8 = 0.796$.

\section{Data and methodology}
\label{sec_data}

\subsection{SDSS observation and galaxy properties}
\label{sec:sdss}

The observational galaxy sample analysed in this work is based on the
New York University Value-Added Galaxy Catalogue (NYU-VAGC)
\cite{Blanton2005}, constructed from the SDSS DR7 spectroscopic data
\cite{York2000,Abazajian2009}. We consider the contiguous Northern
Galactic Cap (NGC) region with
$111^{\circ}<\alpha<264^{\circ}$ and
$-3^{\circ}<\delta<68^{\circ}$, following the survey selection adopted
in \cite{Shi2016}. Within this region, we restrict the galaxy sample to
$0.01\leq z\leq0.12$ and impose the $r$-band absolute-magnitude
criterion
${}^{0.1}M_r-5\log h<-20.09$ to obtain a volume-limited sample.
The absolute magnitudes are $K$-corrected and evolution-corrected to
$z=0.1$ \cite{Blanton2003a,Blanton2003b}. These selections yield
158,840 galaxies. This sample serves as the observational reference
throughout this work and is used both to identify cosmic voids and to
investigate galaxy properties as a function of void-centric distance.

To work in approximately real-space coordinates, we adopt the
reconstructed galaxy catalogue of \cite{Shi2016}, in which
redshift-space distortions are approximately corrected using a
group-based reconstruction method. The details of the reconstruction
procedure are given in \cite{Shi2016} and are not repeated here.
We convert the reconstructed redshifts and angular coordinates into
comoving Cartesian coordinates and use the resulting positions in the
void analysis and in measuring galaxy distances from void centres.

Stellar masses are estimated using the colour-dependent mass-to-light ratio
relation of \cite{Bell2003} and are expressed as
$\log(M_*/h^{-2}M_\odot)$. Galaxy colours are characterized by the
${}^{0.1}(g-r)$ colour, with magnitudes $K$-corrected and
evolution-corrected to $z=0.1$. 

Star formation activity is characterized by sSFR. We use the
public catalogue of \cite{Chang2015}, which combines SDSS and
Wide-field Infrared Survey Explorer (WISE) \cite{Wright2010}
photometry and employs the Multi-wavelength Analysis of Galaxy
Physical Properties (MAGPHYS) \cite{daCunha2008} to model
stellar and dust emission. The catalogue provides star formation
rates with aperture corrections. We use the catalogue's validity
flag to select galaxies with reliable sSFR measurements, leaving
141,701 galaxies.

Galaxy morphology is investigated using the Galaxy Zoo 2 (GZ2)
catalogue \cite{Willett2013}, which provides morphological
classifications for SDSS DR7 galaxies based on volunteer responses
to a multi-step decision tree. We use the \texttt{gz2\_class}
classification, treating galaxies whose class begins with `E' as
elliptical and those beginning with `S' as spiral; galaxies with other
or missing classifications are excluded. This selection yields 105,267
classified galaxies, including 62,111 spirals and 43,156 ellipticals.
These classifications are used to quantify the morphological
composition of galaxies as a function of void-centric distance.

We further characterize galaxy structure using the $r$-band Petrosian
half-light radius, $R_{50}$, from the NYU-VAGC imaging catalogue.
Defined as the radius enclosing 50\% of the Petrosian flux, $R_{50}$
is expressed in $h^{-1}\,\mathrm{kpc}$. 

Together, these five properties allow us to examine the radial variation of galaxy stellar content,
colour, star formation activity, morphological composition, and size
within cosmic voids.

\subsection{ELUCID simulation and subhalo catalogue}
\label{sec:elucid}

The ELUCID simulation is a dark-matter-only constrained simulation
of the local large-scale structure \cite{WangHuiyuan2012,WangHuiyuan2014,
WangHuiyuan2016}. Its initial conditions are constrained by the observed
distribution of nearby galaxy groups, and the simulation is designed
to reproduce the large-scale matter distribution of the local Universe.
In this work, we use ELUCID as a dynamically reconstructed realization
of the nearby large-scale structure and investigate the properties of
dark-matter subhaloes within its reconstructed region.

The simulation starts from $3072^3$ dark-matter particles at
$z=100$ in a periodic box of $500\,h^{-1}\,\mathrm{Mpc}$ on a side.
The particle mass is $3.1\times10^8\,h^{-1}M_\odot$. The density field
is evolved to $z=0$ using a memory-efficient version of the GADGET-2
code \cite{Springel2005}, with 100 available snapshots spanning
$z=18.4$ to $z=0$. Dark-matter haloes are identified using the
Friends-of-Friends (FOF) algorithm \cite{Davis1985}, with a linking
length of $b=0.2$ times the mean inter-particle separation. The FOF
haloes are subsequently decomposed into gravitationally bound subhaloes
using the SUBFIND algorithm \cite{Springel2001}.

Merger trees are constructed by linking subhaloes across successive
snapshots. The formation redshift, $z_{\rm f}$, is defined as the
redshift at which the main progenitor branch reaches half of its
maximum mass, $M_{\rm peak}$ \cite{Zhang2021b}. We characterize
subhaloes using their mass, formation redshift, size, concentration,
and spin. The dimensionless Peebles spin parameter is defined as
\citep{Peebles1969}
\begin{equation}
\label{eqn:spin}
\lambda =
\frac{J|E|^{1/2}}{G M_{\rm sub}^{5/2}},
\end{equation}
where $M_{\rm sub}$, $J$, and $E$ denote the total mass, angular
momentum, and energy of the subhalo, respectively, all calculated
from its member dark-matter particles. The subhalo mass, formation
redshift, spin, and half-mass radius $R_{1/2}$ (the radius enclosing
half of $M_{\rm sub}$) are measured for all subhaloes, including both
central and satellite subhaloes.

The virial radius and concentration characterize the host halo
associated with each central subhalo. For each central subhalo, we
take the position of its minimum gravitational potential as the host
halo centre and use all dark-matter particles around this position.
The virial radius, $R_{\rm vir}$, is defined in a spherical-overdensity
sense as the radius within which the mean enclosed density is
$\Delta_{\rm vir}$ times the critical density of the Universe
\cite{Bryan1998}, and $M_{\rm vir}$ is the mass enclosed within
$R_{\rm vir}$. The virial concentration is defined as
\begin{equation}
\label{eqn:cvir}
c_{\rm vir}=\frac{R_{\rm vir}}{r_{\rm s}},
\end{equation}
where $r_{\rm s}$ is the scale radius obtained by fitting an NFW 
profile \citep{Navarro1997} to the density distribution around this centre.
Thus, $R_{\rm vir}$ and $c_{\rm vir}$ characterize the host halo
associated with a central subhalo, whereas $R_{1/2}$ characterizes the
internal mass distribution of an individual subhalo.

For central subhaloes in the ELUCID simulation, the mean ratio of
subhalo mass to host halo virial mass is $M_{\rm sub}/M_{\rm vir}=0.99\pm0.07$ for
$M_{\rm sub}\geq10^{11.8}\,h^{-1}M_\odot$ and
$1.03\pm0.09$ for $M_{\rm sub}\geq10^{11.0}\,h^{-1}M_\odot$,
where the uncertainties denote the standard deviations across the
respective samples. These ratios indicate that $M_{\rm sub}$ is a
good proxy for host halo mass when dividing central subhaloes into
mass ranges.

For the analysis below, we use subhaloes located within the
reconstructed region corresponding to the SDSS volume-limited sample.

\subsection{Tracer samples}
\label{subsec:subhalo_samples}

We analyse three tracer samples. The first is the SDSS volume-limited
galaxy sample described in Section~\ref{sec:sdss}, which serves as
the observational reference.

The second is an SDSS-matched subhalo sample selected from the ELUCID
simulation within the same volume as the SDSS volume-limited galaxy
sample. We impose a lower mass threshold such that the number of
selected subhaloes equals the number of SDSS galaxies, which yields
$M_{\rm sub}\geq10^{11.8}\,h^{-1}M_\odot$. This sample therefore has
the same number density as the observed galaxy sample and is used to
identify ELUCID voids, in the same way that the SDSS galaxies are used
to identify SDSS voids.

The third is an extended low-mass subhalo sample with
$10^{11.0}\leq M_{\rm sub}<10^{11.8}\,h^{-1}M_\odot$. The lower
mass limit corresponds to approximately 320 dark-matter particles
and is adopted as a practical resolution limit. We measure the
properties of these lower-mass subhaloes within the same ELUCID voids,
which are identified using the
$M_{\rm sub}\geq10^{11.8}\,h^{-1}M_\odot$ sample. Comparing the two
subhalo samples allows us to investigate how the radial dependence
of subhalo properties varies with subhalo mass.

\subsection{Void catalogues}
\label{subsec:voids}

\begin{figure*}[htbp]
\centering
\includegraphics[width=1.0\textwidth]{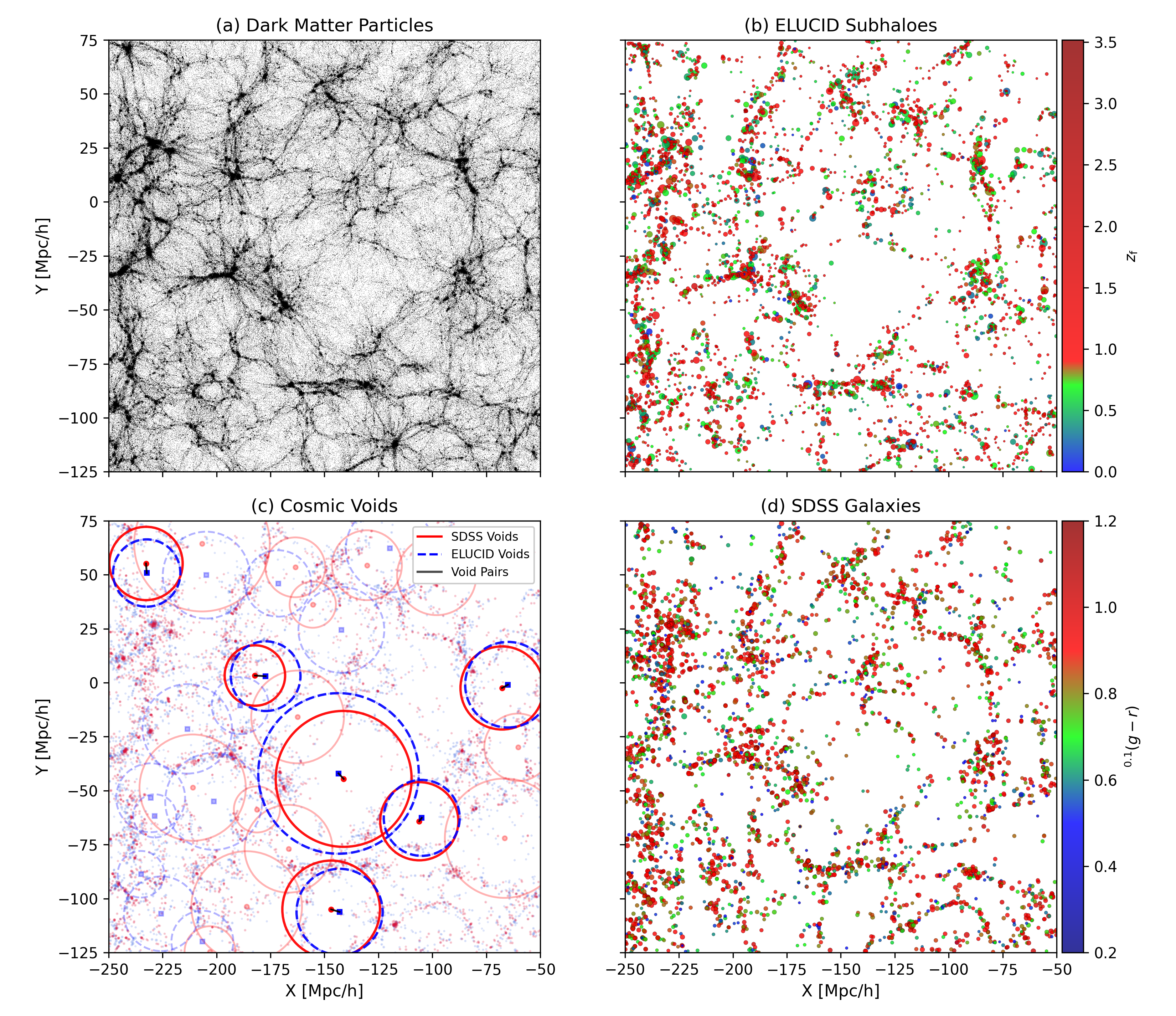}
\caption{
Spatial distributions of the matter, subhalo, void, and galaxy
samples used in this work within a $15\,h^{-1}\mathrm{Mpc}$-thick
slice. Panel~(a) shows a $1\%$ random subsample of dark-matter
particles from the ELUCID simulation. Panel~(b) shows ELUCID
subhaloes with $M_{\rm sub}\geq10^{11.8}\,h^{-1}M_\odot$, with marker
size indicating subhalo mass and colour indicating formation redshift.
Panel~(c) shows the SDSS and ELUCID void catalogues, with thick
circles and connecting lines highlighting the 102 corresponding
SDSS--ELUCID void pairs. Panel~(d) shows SDSS galaxies in the
volume-limited sample, with marker size indicating stellar mass and
colour indicating galaxy colour.
}
\label{fig:void_slice}
\end{figure*}

Cosmic voids are identified independently in the SDSS galaxy sample
and the SDSS-matched ELUCID subhalo sample
(Section~\ref{subsec:subhalo_samples}) using the \texttt{REVOLVER}
pruning mode \citep{Nadathur2019}, implemented within the \texttt{VAST}
framework \citep{Douglass2023}. The voids are defined as watershed
basins constructed from the Voronoi tessellation of the tracer
distribution.

For each void, we use the volume-weighted centre,
\begin{equation}
\label{eqn:center}
\mathbf{C} =
\frac{\sum_i \mathbf{x}_i V_i}
     {\sum_i V_i},
\end{equation}
where $\mathbf{x}_i$ and $V_i$ are the position and Voronoi cell
volume of the $i$-th tracer in the void. The effective void radius is
defined as
\begin{equation}
\label{eqn:void_radius}
R_{\rm v} =
\left(
\frac{3V_{\rm void}}{4\pi}
\right)^{1/3},
\end{equation}
where $V_{\rm void}$ is the total watershed volume. The void centre
and effective radius defined above are used to calculate the normalized
void-centric radius $r/R_{\rm v}$ throughout the radial analysis.
We use $r/R_{\rm v}=1$ as the reference location corresponding to the
effective void radius when describing radial trends relative to the
void boundary. This location does not necessarily coincide with the
actual boundary of an individual void, whose watershed region can have
an irregular, non-spherical shape.

For the SDSS catalogue, voids are identified from the reconstructed
real-space galaxy sample described in Section~\ref{sec:sdss}. For the
ELUCID catalogue, the same void-finding procedure is applied to the
SDSS-matched subhalo sample, which has the same number density as the
SDSS galaxies. We adopt the same \texttt{REVOLVER} definition for both
datasets to ensure a
consistent void-identification procedure. In our previous comparison
of SDSS and ELUCID voids, \texttt{REVOLVER} yielded more consistent
radial number-density measurements between the two tracer populations
than the geometry-based \texttt{VoidFinder} method
\citep{Zhang2026a}. 

We initially identify voids down to $R_{\rm v}=5\,h^{-1}\mathrm{Mpc}$
and retain those with $R_{\rm v}\geq10\,h^{-1}\mathrm{Mpc}$ for the
subsequent analysis. To exclude voids significantly affected by survey
or reconstruction boundaries, we use the total surface area of each
void, $A_{\rm total}$, and the part of that surface shared with the
Voronoi cells of boundary tracers, $A_{\rm edge}$, both provided in the
\texttt{VAST} void catalogues \citep{Douglass2023}. We require
$A_{\rm edge}/A_{\rm total}<0.1$. These selections yield 389 SDSS voids
and 404 ELUCID voids.

To identify corresponding voids in the two catalogues, we apply the
matching scheme adopted in \citep{Zhang2026b}. Candidate pairs are
evaluated according to their geometric overlap, quantified by the
intersection-over-union (IoU) of equivalent spheres with radii given
by their effective void radii. We retain candidate pairs with
${\rm IoU}>0.4$ and accept a pair only when each void identifies the
other as its highest-IoU counterpart. This mutual-best criterion
yields 102 corresponding SDSS--ELUCID void pairs.

The resulting spatial distributions of the tracer and void samples
are shown in Figure~\ref{fig:void_slice}. The SDSS galaxies in
panel~(d) and the ELUCID subhaloes in panel~(b) trace the same
large-scale structures as the dark-matter distribution in panel~(a),
including the large void near
$(X,Y)\approx(-142,-43)\,h^{-1}\mathrm{Mpc}$, as expected for a
constrained simulation. In the galaxy distribution, redder galaxies
are preferentially concentrated around dense nodes and filaments.
Panel~(c) shows that the 102 corresponding SDSS--ELUCID void pairs 
form only a subset of the full catalogues, representing approximately
one quarter of the 389 SDSS voids and the 404 ELUCID voids.

For all galaxy and subhalo properties considered in this work, the
radial profiles measured using the corresponding SDSS--ELUCID void
pairs are broadly consistent with those obtained from the full
catalogues, although the uncertainties are larger for the smaller
sample. As representative examples, we show the comparisons
for the stellar mass and subhalo mass profiles in
Figures~\ref{fig:void_stellar_mass} and \ref{fig:void_subhalo_mass}.
Given this consistency and the smaller uncertainties afforded by the
larger samples, we adopt the full SDSS and ELUCID void catalogues for
the radial measurements presented below. 

\subsection{Statistical measurements}
\label{subsec:statistics}

Throughout this work, the radial dependence of galaxy and subhalo
properties is measured as a function of the normalized void-centric
radius $r/R_{\rm v}$, using 16 equal-width radial bins in the range
$0\leq r/R_{\rm v}<2.5$. The measurements are performed on a
void-by-void basis. For each void and each radial bin, we first
calculate the mean value of the property of interest from the galaxies
or subhaloes within that bin. We then average these void-level
measurements over all contributing voids. Thus, each void contributes
equally to the final measurement, rather than weighting the result by
the number of tracers it contains. Since neighbouring voids can have
overlapping regions, the same tracer can contribute to the measurements
of more than one void.

A void contributes to a radial bin if that bin contains at least one
tracer, and we require each radial bin to be contributed by at least
10 voids; bins with fewer contributing voids are excluded. The
statistical uncertainty in each radial bin is estimated from the
standard error of the mean of the void-level measurements. This
procedure is applied consistently to the galaxy and subhalo
measurements throughout the paper.

We adopt a minimum of one tracer per void and radial bin because a
larger threshold would preferentially remove the tracer-poor interiors
of voids and could introduce a selection effect that depends on tracer
abundance. We have also tested minimum tracer numbers ranging from 2 to
10. The resulting radial profiles are consistent with those presented
here, but the innermost radial bins are contributed by progressively
fewer voids as the threshold increases, so that more of them fall below
the 10-void requirement and are removed. A threshold of one tracer
therefore retains the most information on the void interiors.

\section{Results}
\label{sec_results}

\subsection{Galaxy properties in SDSS voids}
\label{sec:galaxy_properties}

We investigate the dependence of several galaxy properties on
void-centric distance, including stellar mass, galaxy colour, sSFR, 
morphology, and galaxy size. Unless otherwise
stated, all measurements are performed using the volume-limited SDSS
galaxy sample described in Section~\ref{sec:sdss}.

To separate the dependence on stellar mass from the residual
dependence on void-centric distance, we further divide the
volume-limited sample into three stellar-mass subsamples with
approximately equal numbers of galaxies in the full volume-limited sample.
The three mass ranges are
$9.6 \leq \log(M_*/h^{-2}M_\odot) < 10.4$,
$10.4 \leq \log(M_*/h^{-2}M_\odot) < 10.6$, and
$\log(M_*/h^{-2}M_\odot) \geq 10.6$.
This division is adopted consistently when investigating the dependence
of galaxy colour, sSFR, morphology, and size on
void-centric distance. In addition to the mass-selected subsamples, the
full volume-limited sample is used to characterize the overall radial
trend of each galaxy property.

\subsubsection{Stellar mass}
\label{sec_mass}

\begin{figure}[htbp]
  \centering
  \includegraphics[width=0.8\textwidth]{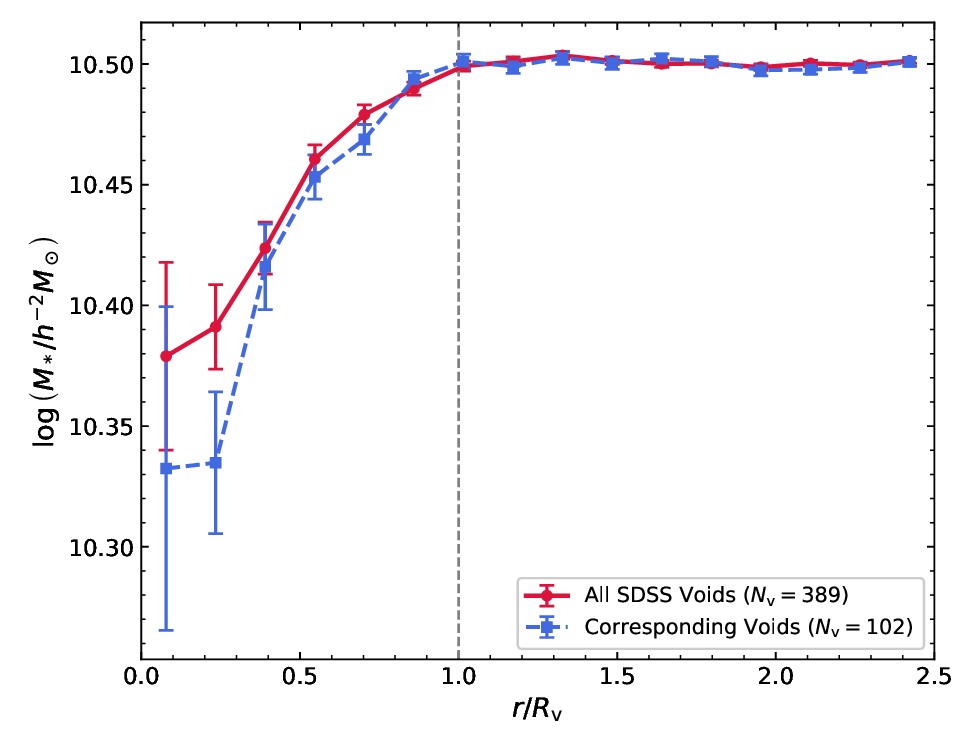}
  \caption{Radial dependence of the mean stellar mass of the
  volume-limited galaxy sample as a function of void-centric distance,
  normalized by the effective void radius, $r/R_{\rm v}$. The red solid
  curve shows the result for all SDSS voids ($N_{\rm v}=389$), while the
  blue dashed curve shows the measurement for the 102 SDSS voids that
  have a corresponding void in the ELUCID catalogue. Error bars denote
  the standard error of the mean estimated from the void-to-void scatter.
  The vertical dashed line marks the void boundary at $r/R_{\rm v}=1$.
  Only radial bins contributed by at least 10 voids are displayed.}
  \label{fig:void_stellar_mass}
\end{figure}

Figure~\ref{fig:void_stellar_mass} shows the radial dependence of the
mean stellar mass of the volume-limited galaxy sample around SDSS void
centres. The same sample, defined by
${}^{0.1}M_r - 5\log h < -20.09$, is used both to identify the voids and
to characterize the galaxies residing within and around them.

A clear radial dependence of the mean stellar mass is found within the
void interiors. The galaxies near the void centres have the lowest mean
stellar masses, with $\log(M_*/h^{-2}M_\odot)\simeq10.38$ at
$r/R_{\rm v}\simeq0.1$. The mean stellar mass then increases steadily
with void-centric distance, most rapidly at
$0.2\lesssim r/R_{\rm v}\lesssim0.6$, and reaches
$\log(M_*/h^{-2}M_\odot)\simeq10.50$ close to the void boundary. The
total increase from the innermost radial bin to $r/R_{\rm v}\simeq1$ is
about $0.12$ dex, indicating a systematic increase in the typical
stellar mass of galaxies from void centres toward their boundaries.
Outside the void boundary, the mean stellar mass remains close to
$\log(M_*/h^{-2}M_\odot)\simeq10.50$ out to $r/R_{\rm v}=2.5$, with only
minor fluctuations. The dependence of galaxy stellar mass on void-centric
distance is therefore primarily confined to the void interior.

This result is in line with previous findings that galaxies in the
inner regions of voids tend to have lower stellar masses than those in
denser environments. In the Horizon-AGN simulation, the mean stellar
mass of galaxies increases with void-centric distance in the inner
regions of voids and flattens around the void boundary
\cite{Habouzit2020}. In a semi-analytic model applied to the
Millennium simulation in a Euclid-like lightcone, the stellar-mass
distributions of galaxies in bins of void-centric distance shift to
lower masses closer to void centres, although the differences between
bins are small \cite{Euclid2026}. In the EAGLE simulation, galaxies
were divided into inner-void, outer-void, wall, and skeleton samples
according to their void-centric distance, and inner-void galaxies were
found to have lower stellar masses than those in denser environments
\cite{Rosas-Guevara2022}. These studies are based on hydrodynamical simulations or simulated
catalogues. Complementing them, the present measurement shows that a
consistent trend, including its flattening near the void boundary, is
also seen in observed SDSS galaxies as a function of $r/R_{\rm v}$.

The blue dashed curve in Figure~\ref{fig:void_stellar_mass} shows the
result for the 102 SDSS voids with a corresponding ELUCID void. It
follows the same overall trend as the full sample, with the two
measurements nearly indistinguishable for $r/R_{\rm v}\gtrsim0.5$. The
slightly lower stellar masses in the innermost bins are accompanied by
large statistical uncertainties.

\subsubsection{Galaxy colour}
\label{sec_color}

\begin{figure}[htbp]
  \centering
  \includegraphics[width=0.8\textwidth]{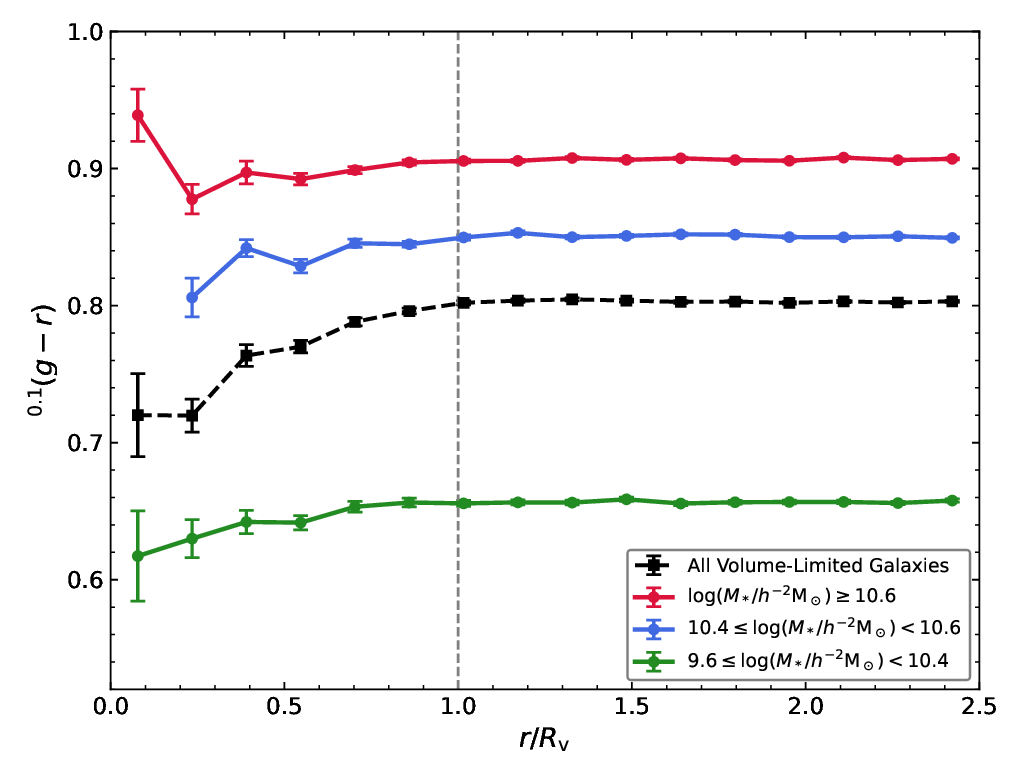}
  \caption{Radial dependence of the mean galaxy $^{0.1}(g-r)$ colour around SDSS cosmic voids, as a function of the normalized void-centric distance
  $r/R_{\rm v}$. The black curve represents the full volume-limited galaxy sample, while the coloured curves show three stellar-mass
  subsamples containing approximately equal numbers of galaxies. The vertical dashed line indicates the effective
  void boundary, $r/R_{\rm v}=1$. Only radial bins contributed by at least 10 voids are displayed.}
  \label{fig:void_color}
\end{figure}

Figure~\ref{fig:void_color} shows the radial dependence of the mean
${}^{0.1}(g-r)$ colour for the full volume-limited galaxy sample and for
the three stellar-mass subsamples defined in
Section~\ref{sec:galaxy_properties}. The full sample exhibits a clear
increase in mean colour toward the void boundary. At
$r/R_{\rm v}\simeq0.1$ the mean colour is about
${}^{0.1}(g-r)\simeq0.72$, increasing to approximately $0.80$ at the
void boundary, and it remains nearly constant at
${}^{0.1}(g-r)\simeq0.80$ outside the voids.

As shown in Figure~\ref{fig:void_color}, the colour depends strongly on stellar mass. 
The most massive galaxies, $\log(M_*/h^{-2}M_\odot)\geq10.6$, 
are systematically redder than the
lower-mass populations at all void-centric distances, with a mean colour
of $\simeq0.91$ near the void boundary. The lowest-mass galaxies,
$9.6\leq\log(M_*/h^{-2}M_\odot)<10.4$, are substantially bluer, with
mean colours increasing from about $0.62$ near the void centre to
$\simeq0.66$ near the void boundary. The intermediate-mass population
lies in between, increasing from about $0.81$ at
$r/R_{\rm v}\simeq0.2$ to $\simeq0.85$ at the boundary. For the most
massive galaxies, the colour varies by less than about $0.03$ at
$r/R_{\rm v}\gtrsim0.2$ without a clear monotonic trend; the innermost
point is redder, but it has a large uncertainty. For all three mass
ranges the profiles become nearly flat at $r/R_{\rm v}\gtrsim1$.

The radial variation of colour within each mass subsample
($\lesssim0.05$) is therefore smaller than the change of about $0.08$
seen in the full sample, and much smaller than the difference between
the mass subsamples, which reaches $\simeq0.25$ between the lowest- and
highest-mass populations at the void boundary. Since the mean stellar
mass of galaxies increases toward the void boundary
(Section~\ref{sec_mass}), part of the colour gradient in the full sample
reflects the changing mass composition. A residual trend toward bluer
colours in void interiors nevertheless remains at fixed stellar mass,
most clearly for the lower- and intermediate-mass galaxies.

The bluer colours of galaxies in void interiors are consistent with
earlier findings that galaxies in voids are on average bluer than those
in denser environments \cite{Rojas2004,Hoyle2012,Florez2021}, and with
the reddening of galaxies from voids toward filaments reported by
\cite{Kuutma2017}. \cite{Hoyle2012} found that red and blue void
galaxies have similar distributions of void-centric distance. 
Our measurement of the mean colour as a function of
$r/R_{\rm v}$ shows a gradient in the full sample, which is partly
related to the stellar-mass segregation, and a weaker residual gradient
at fixed stellar mass.

\subsubsection{Specific star formation rate}
\label{sec_ssfr}

\begin{figure}[htbp]
  \centering
  \includegraphics[width=0.8\textwidth]{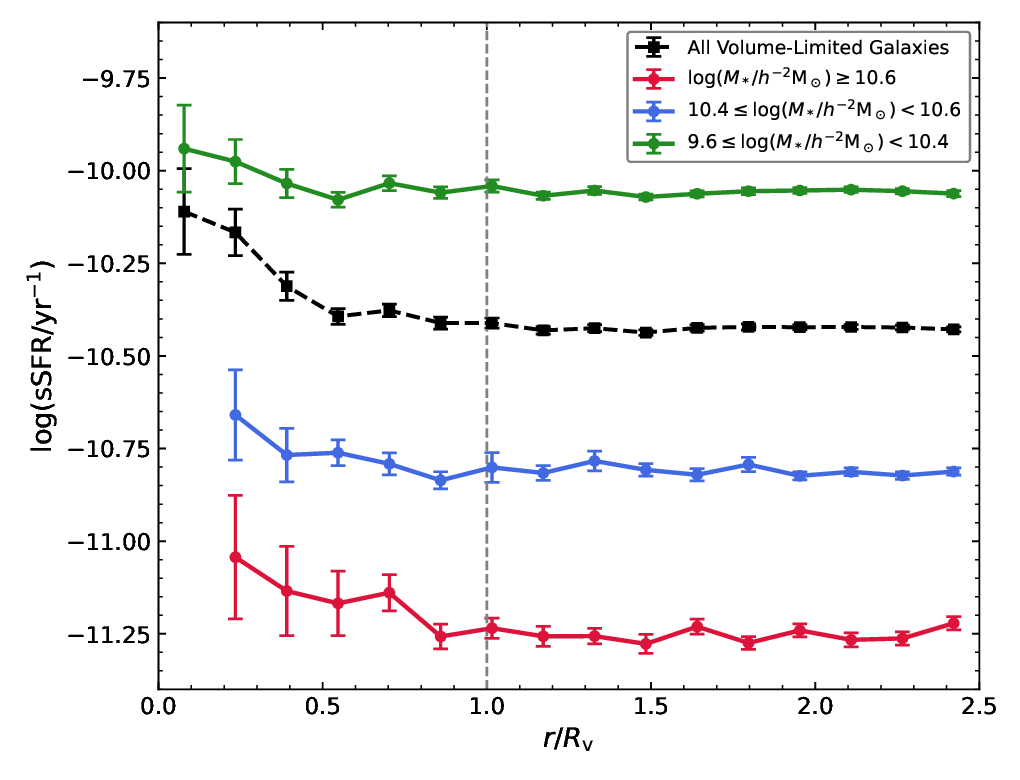}
  \caption{Mean specific star formation rate (sSFR) as a function of
  void-centric distance for the full volume-limited sample and three
  stellar-mass subsamples. The mean sSFR is calculated in linear space
  before taking the logarithm. The dashed vertical line marks
  $r/R_{\rm v}=1$; error bars denote the standard error of the mean
  across voids. Only radial bins contributed by at least 10 voids are
  displayed.}
  \label{fig:void_ssfr}
\end{figure}

Figure~\ref{fig:void_ssfr} presents the radial dependence of the sSFR
taken from \cite{Chang2015} (Section~\ref{sec:sdss}), for the full
volume-limited sample and the three stellar-mass subsamples defined in
Section~\ref{sec:galaxy_properties}. To mitigate the impact of the
extremely low-sSFR tail, the mean sSFR in each radial bin is first
computed in linear space and then converted to
$\log (\mathrm{sSFR}/\mathrm{yr}^{-1})$. The mean is therefore
dominated by star-forming galaxies, and it reflects both the fraction
of star-forming galaxies and their typical sSFR.

For the full volume-limited sample, the mean sSFR decreases with
increasing void-centric distance in the void interior, from
$\log(\mathrm{sSFR}/\mathrm{yr}^{-1})\simeq-10.1$ in the innermost
bin to $\simeq-10.4$ at $r/R_{\rm v}\simeq1$. The decline is most
pronounced at $r/R_{\rm v}\lesssim0.6$, beyond which the profile is
nearly flat, and the mean sSFR remains approximately constant out to
$r/R_{\rm v}\simeq2.5$.

The sSFR depends strongly on stellar mass. At large void-centric
distances, the low-, intermediate-, and high-mass subsamples have
$\log(\mathrm{sSFR}/\mathrm{yr}^{-1})\simeq-10.1$, $-10.8$, and
$-11.2$, respectively. Within each subsample, the sSFR decreases
modestly toward the void boundary, by about $0.1$, $0.15$, and
$0.2$~dex for the low-, intermediate-, and high-mass galaxies, although
the innermost bins have large uncertainties (the intermediate- and
high-mass curves do not extend to the innermost bin owing to the
requirement of at least 10 contributing voids). These changes are
smaller than the $\simeq0.3$~dex decline of the full sample and much
smaller than the difference of about $1.1$~dex between the lowest- and
highest-mass subsamples. A part of the radial decline in the full
sample therefore reflects the changing stellar-mass composition, but a
small residual trend in the same direction remains at fixed stellar
mass.

Higher sSFRs for galaxies in voids than for those in walls have been reported 
by \cite{Rojas2005, Moorman2016, Florez2021}. \cite{Rojas2005} attributed this 
difference partly to the lower stellar masses of void galaxies, 
while \cite{Moorman2016} and \cite{Florez2021} found enhanced sSFRs at
fixed stellar mass and fixed baryonic mass, respectively.
For a sample of 59 void galaxies of lower stellar masses, \cite{Beygu2016}
found that the sSFR decreases with stellar mass in the same way as for
galaxies in denser environments, and is only slightly elevated relative
to field galaxies of the same stellar mass. Our results are compatible
with this picture: the sSFR in void interiors is enhanced partly
through the stellar-mass composition, while the residual dependence at
fixed stellar mass is small compared with the dependence on stellar
mass itself. The decrease of the star formation activity from voids
toward filaments was reported by \cite{Kuutma2017}. In the
Horizon-AGN simulation, the sSFR of galaxies slightly increases toward
void centres, and the authors find that this cannot be attributed only
to the prevalence of low-mass galaxies \cite{Habouzit2020}.

Together with the bluer colours in Section~\ref{sec_color}, these
results indicate that galaxies in void interiors have enhanced star
formation activity, which is partly accounted for by their lower
stellar masses, while the dependence on stellar mass remains much
stronger than the dependence on void-centric distance.

\subsubsection{Galaxy morphology}
\label{sec:morphology}

\begin{figure}[htbp]
    \centering
    \includegraphics[width=0.8\textwidth]{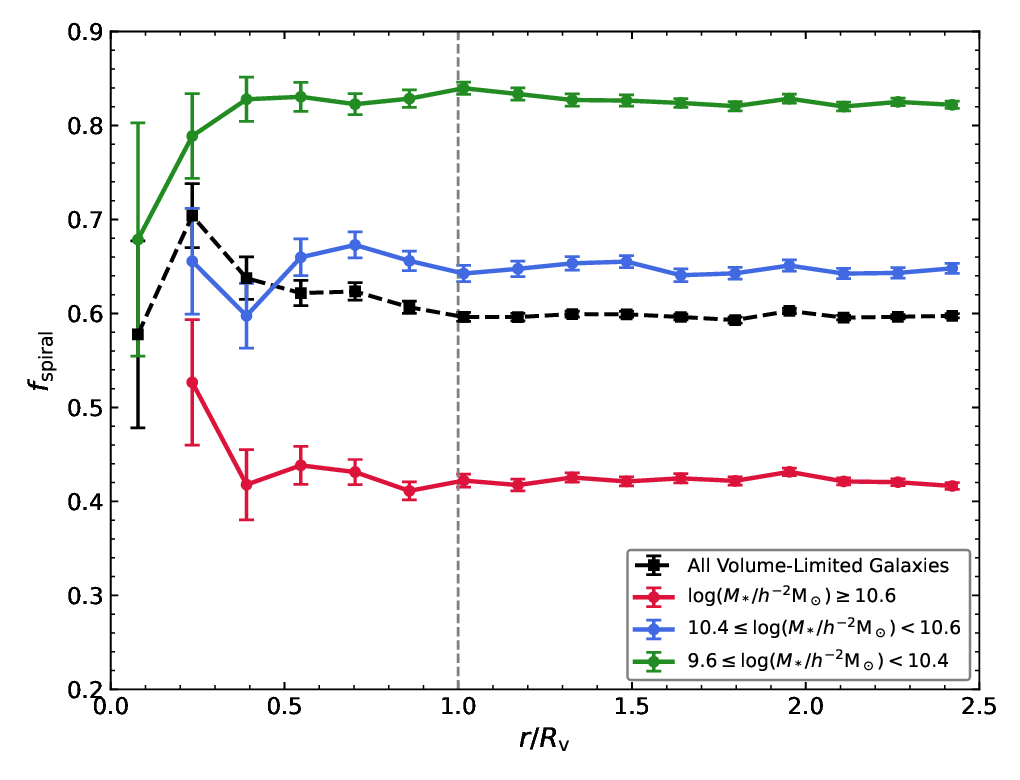}
    \caption{Spiral-galaxy fraction as a function of void-centric distance
    for the full volume-limited sample and three stellar-mass subsamples.
    The spiral fraction is defined as
    $f_{\rm spiral}=N_{\rm spiral}/(N_{\rm spiral}+N_{\rm elliptical})$.
    For each void and radial bin, the fraction is first calculated from
    the galaxies within that bin and then averaged over voids. Error bars
    denote the standard error of the mean across voids. The dashed line
    marks $r/R_{\rm v}=1$. Only radial bins contributed by at least 10
    voids are shown.}
    \label{fig:void_morphology}
\end{figure}

Figure~\ref{fig:void_morphology} shows the radial dependence of the
spiral fraction for the full volume-limited sample and the three
stellar-mass subsamples, using the GZ2 classifications described in
Section~\ref{sec:sdss}. For each void and radial bin, the fraction is
first calculated from the classified galaxies in that bin and then
averaged over voids, giving equal weight to individual voids.

For the full volume-limited sample, the spiral fraction decreases
weakly from $\sim0.64$ at $r/R_{\rm v}\sim0.4$ to $\sim0.60$ around
$r/R_{\rm v}\sim1$, and remains approximately constant at larger
distances. The innermost bins ($r/R_{\rm v}\lesssim0.3$) have large
uncertainties and are not used to characterize the trend. The three
stellar-mass subsamples have very different spiral fractions of
approximately $0.42$, $0.65$, and $0.82$ for the high-, intermediate-,
and low-mass populations, respectively, but within each of them the
fraction varies by less than about $0.05$ at $r/R_{\rm v}\gtrsim0.4$,
with no systematic radial trend within the uncertainties.

Thus, at fixed stellar mass, galaxy morphology shows little systematic
dependence on void-centric distance. The decrease of about $0.04$ in the
full sample is much smaller than the difference of about $0.2$ between
adjacent mass subsamples. Weighting the spiral fractions of the three
subsamples by their relative numbers in each radial bin reproduces the
decline of the full sample within the uncertainties, indicating that the 
shift of the stellar-mass composition toward lower masses in void 
interiors (Section~\ref{sec_mass}) contributes substantially to the
trend of the full sample.

Our radial measurements are consistent with \cite{Argudo2024}, who found that the 
fractions of early- and late-type galaxies remain nearly constant with 
void-centric distance within SDSS-DR7 voids, with a slightly higher 
late-type fraction at $0.1\lesssim r/R_{\rm v}\lesssim0.6$. 
\cite{Ricciardelli2017} reported a declining spiral-galaxy fraction with 
increasing void-centric distance, based on 
volume-limited samples extending to lower stellar masses than ours.
\cite{Florez2021} found more late types among void galaxies, as expected
for their lower baryonic masses. A similar role of galaxy population composition
has been suggested for colour \cite{Bruton2020}. In the EAGLE simulation, \cite{Rosas-Guevara2022} 
found that the fraction of star-forming galaxies decreases with void-centric
distance, although at fixed stellar mass, inner-void galaxies have star 
formation activity comparable to that of galaxies in denser environments 
in some mass ranges.

\subsubsection{Galaxy size}
\label{sec:galaxy_size}

\begin{figure}[htbp]
    \centering
    \includegraphics[width=1.0\textwidth]{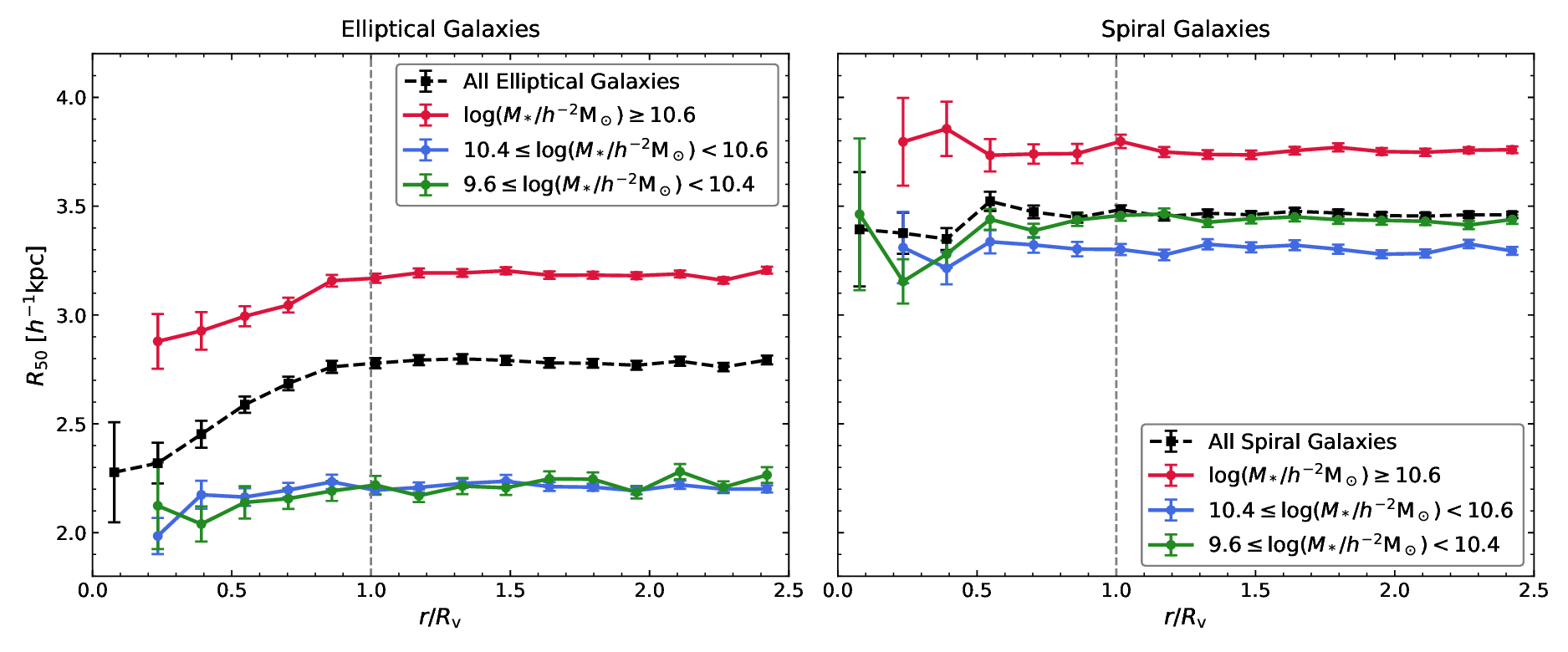}
    \caption{Mean galaxy half-light radius $R_{50}$ as a function of
    void-centric distance for elliptical (left) and spiral (right)
    galaxies. Error bars denote the standard error of the mean across
    voids. The vertical dashed line marks the void effective radius,
    $r/R_{\rm v}=1$. Only radial bins contributed by at least 10 voids
    are shown.}
    \label{fig:void_r50}
\end{figure}

Given the strong dependence of galaxy size on morphology, we measure
$R_{50}$ separately for elliptical and spiral galaxies, using the same
three stellar-mass subsamples defined in
Section~\ref{sec:galaxy_properties}. Figure~\ref{fig:void_r50} shows the
radial dependence of $R_{50}$, with the black dashed curves for the
full morphological samples and the coloured curves for the stellar-mass
subsamples. 

For elliptical galaxies, the mean $R_{50}$ of the full sample increases
from about $2.3\,h^{-1}$kpc in the inner void regions to about
$2.8\,h^{-1}$kpc at $r/R_{\rm v}\sim1$, and is approximately constant at
larger radii. The high-mass ellipticals, whose sizes
($\simeq3.2\,h^{-1}$kpc) are about $1\,h^{-1}$kpc larger than those of
the lower-mass ellipticals, show an increase of about $10\%$ over the
same range, while the intermediate- and low-mass ellipticals show only a
weak increase of $\lesssim0.2\,h^{-1}$kpc, comparable to the
uncertainties, with mean sizes of about $2.2\,h^{-1}$kpc at larger
radii. The increase of the full elliptical sample is therefore larger
than that within any of the mass subsamples, and is partly related to
the shift of the stellar-mass composition toward lower masses in void
interiors (Section~\ref{sec_mass}).

The high-mass subsample ($\log(M_*/h^{-2}M_\odot)\geq10.6$) spans about
$1.1$ dex in stellar mass, up to $\log(M_*/h^{-2}M_\odot)\simeq11.7$,
and the size of elliptical galaxies depends strongly on stellar mass: the
mean $R_{50}$ rises from about $2.5\,h^{-1}$kpc at the low-mass end of
this range to above $6\,h^{-1}$kpc at the high-mass end. A change of the
mean stellar mass within the bin with void-centric distance can therefore
produce an apparent radial trend of the mean size. To test this, we
divide the high-mass ellipticals into ten bins of width
$0.11$ dex in stellar mass. Within these narrower bins, the radial
dependence of $R_{50}$ is largely absent at $r/R_{\rm v}\gtrsim0.5$,
with variations within the uncertainties; the most massive bins contain
few galaxies and do not extend to small $r/R_{\rm v}$. The increase of
the mean size of the high-mass ellipticals toward the void boundary is
therefore primarily a consequence of mass mixing within the broad bin,
rather than evidence for an independent dependence of galaxy size on
void-centric distance.

For spiral galaxies, the radial dependence of $R_{50}$ is weaker. The
full sample is nearly constant at about $3.4$--$3.5\,h^{-1}$kpc, and the
high-mass spirals remain nearly constant at about
$3.8\,h^{-1}$kpc. The intermediate- and low-mass spirals show modest
variations of about $0.2$--$0.3\,h^{-1}$kpc at $r/R_{\rm v}\lesssim0.5$,
with large uncertainties. Beyond the void boundary, the sizes of all
subsamples are approximately constant. Overall, the radial variation of
the galaxy size is therefore more evident for elliptical galaxies,
particularly at high stellar mass, whereas spiral galaxies show a weak
dependence on void-centric distance.

Previous studies have reported mixed results regarding the environmental
dependence of galaxy sizes. The galaxies in the Void Galaxy Survey have 
small half-light radii, and \cite{Beygu2017} highlighted their low stellar
masses and small sizes as characteristic of the void galaxy population.
For isolated quiescent galaxies, \cite{Mosleh2018} found that void 
galaxies have at most $10$--$25\%$ smaller half-mass and half-light 
sizes than their counterparts in groups and clusters at fixed stellar mass.
In contrast, analyses of the SDSS found no significant dependence of the
size--mass relation on environment at fixed galaxy type, based on comparisons
between cosmic-web environments in SDSS DR7 \cite{Zhang2019} and between
clusters and voids in SDSS DR13 \cite{Abdullah2026}. More recently, \cite{Curtis2026} reported
that void galaxies with $M_*>10^{11}h^{-1}M_\odot$ have half-light 
radii that are $11\pm3\%$ smaller than those of field galaxies at fixed
stellar mass, redshift, and colour. This size deficit becomes stronger
toward the most underdense regions and is not significant at lower stellar masses.

In our sample, galaxy sizes show only weak radial variations within fixed stellar-mass ranges,
and the apparent increase in the mean size of high-mass ellipticals toward
the void boundary largely disappears when narrower stellar-mass bins are used. 
These results suggest that the apparent radial trends in galaxy sizes are 
largely driven by variations in stellar-mass composition, particularly for 
high-mass ellipticals, although they do not directly constrain size offsets
between void and field galaxies.

\subsection{Halo properties in ELUCID voids}
\label{sec:halo_properties}

We next investigate the radial dependence of dark-matter subhalo
properties in ELUCID voids, using the SDSS-matched subhalo sample
($M_{\rm sub}\geq10^{11.8}\,h^{-1}M_\odot$) and the extended low-mass
sample ($10^{11.0}\leq M_{\rm sub}<10^{11.8}\,h^{-1}M_\odot$) described
in Section~\ref{subsec:subhalo_samples}. Both samples are measured
within the same ELUCID void catalogue, so that the void population and
its geometric definition are kept fixed when comparing the two mass
ranges. For both samples we investigate the radial dependence of
subhalo mass, satellite fraction, formation redshift, size, concentration, and spin.

As for the galaxy properties, the formation redshift, size,
concentration, and spin are measured for each full sample and within
narrower subhalo-mass ranges, so that radial trends driven by mass can
be distinguished from those that remain at fixed mass. The SDSS-matched
sample is divided into three subsamples with approximately equal numbers, with
the mass boundaries determined from the full spatially selected sample,
and kept fixed throughout the analysis.
The extended low-mass sample is divided into three fixed intervals,
$11.0\leq\log(M_{\rm sub}/h^{-1}M_\odot)<11.2$,
$11.2\leq\log(M_{\rm sub}/h^{-1}M_\odot)<11.5$, and
$11.5\leq\log(M_{\rm sub}/h^{-1}M_\odot)<11.8$, which provide a more
even coverage of the extended range than an equal-number division.

\subsubsection{Subhalo mass}
\label{sec:subhalo_mass}

\begin{figure}[htbp]
    \centering
    \includegraphics[width=1.0\textwidth]{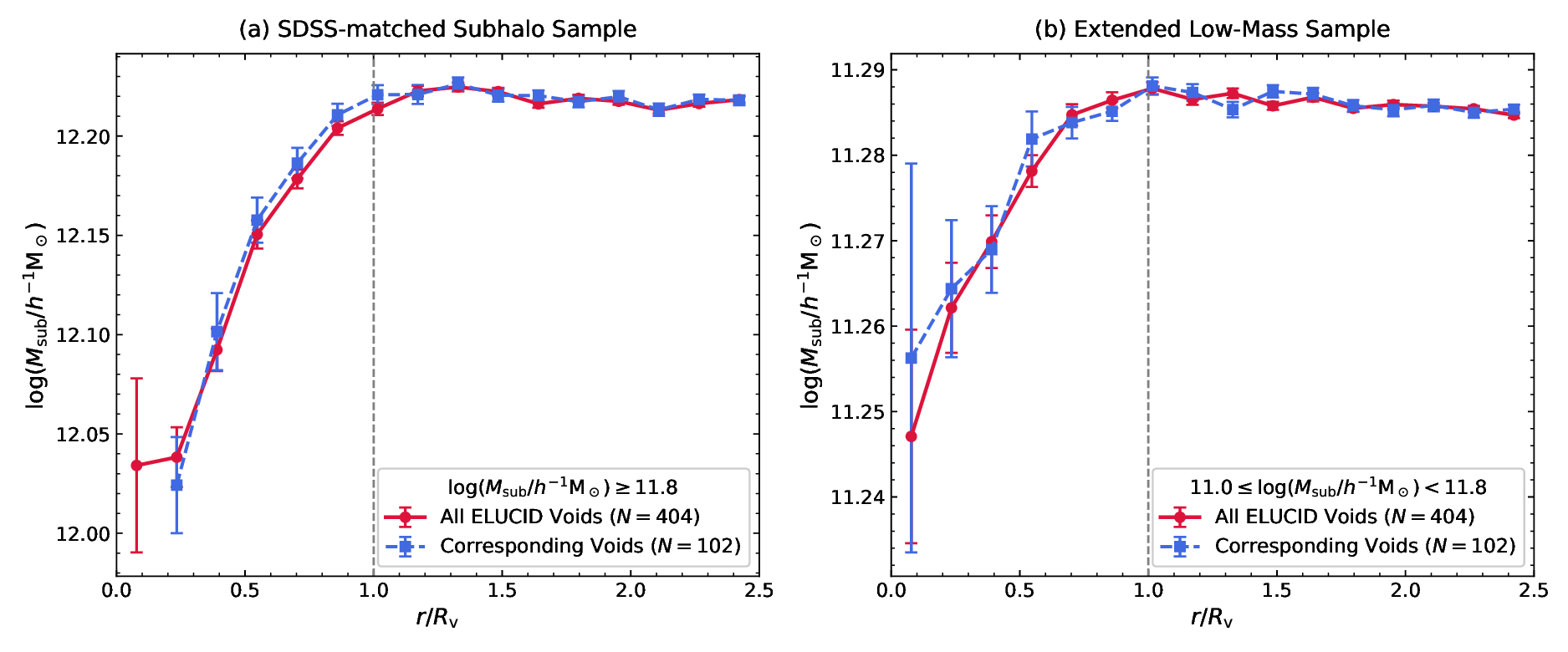}
    \caption{Radial dependence of the mean logarithmic subhalo mass.
    Panel~(a) shows the SDSS-matched sample with
    $M_{\rm sub}\geq10^{11.8}\,h^{-1}M_\odot$, while panel~(b) shows
    the extended low-mass sample with
    $10^{11.0}\leq M_{\rm sub}<10^{11.8}\,h^{-1}M_\odot$.
    Red solid and blue dashed curves show all ELUCID voids
    ($N_{\rm v}=404$) and the 102 ELUCID voids with a counterpart in 
    the SDSS catalogue. 
    Error bars denote the standard error of the mean
    from the void-to-void scatter. The vertical dashed line marks
    $r/R_{\rm v}=1$. Only radial bins contributed by at least 10 voids
    are shown.}
    \label{fig:void_subhalo_mass}
\end{figure}

Figure~\ref{fig:void_subhalo_mass} shows the radial dependence of the
mean logarithmic subhalo mass for the two subhalo samples. For the
SDSS-matched sample shown in panel~(a), the mean subhalo mass increases
systematically with void-centric distance inside the voids. It rises
from approximately $\log(M_{\rm sub}/h^{-1}M_\odot)\simeq12.04$ at $r/R_{\rm v}\simeq0.2$ 
to $\simeq12.22$ near the void boundary,
corresponding to an increase of about $0.18$ dex. Most of this
variation occurs over $0.3\lesssim r/R_{\rm v}\lesssim 1.0$. Beyond the
void boundary, the mean subhalo mass changes only weakly and remains
close to $12.22$ out to $r/R_{\rm v}=2.5$.

A similar radial trend is found for the extended low-mass sample in
panel~(b), although with a substantially smaller amplitude. The mean
subhalo mass increases from approximately
$\log(M_{\rm sub}/h^{-1}M_\odot)\simeq11.25$ near the void centre to
$\simeq11.29$ around $r/R_{\rm v}\simeq1$, a variation of about $0.04$
dex, and remains nearly constant beyond the void boundary. Thus, the
mean subhalo mass increases toward the outskirts of the voids in both
mass ranges, while the amplitude of the radial dependence is stronger
for the higher-mass sample. The results for all ELUCID voids and for the
102 corresponding voids are consistent in both panels, with larger
uncertainties for the smaller sample.

This mass segregation of subhaloes is similar to that of the SDSS
galaxies in Section~\ref{sec_mass}. In N-body simulations, massive haloes
tend to form closer to the void boundaries \cite{Gottlober2003}, and the
halo mass function inside voids depends on the distance from the void
centre \cite{Verza2022}. More specifically, \cite{Alfaro2025} found that
lower-mass haloes tend to populate the inner regions of voids, whereas
more massive haloes are preferentially located near the void walls.
Our results show a similar radial trend in the mean subhalo mass in
the ELUCID simulation, using voids that can be directly compared with
those identified in the SDSS galaxy sample.

\subsubsection{Satellite fraction}
\label{sec:satellite_fraction}

\begin{figure}[htbp]
    \centering
    \includegraphics[width=1.0\textwidth]{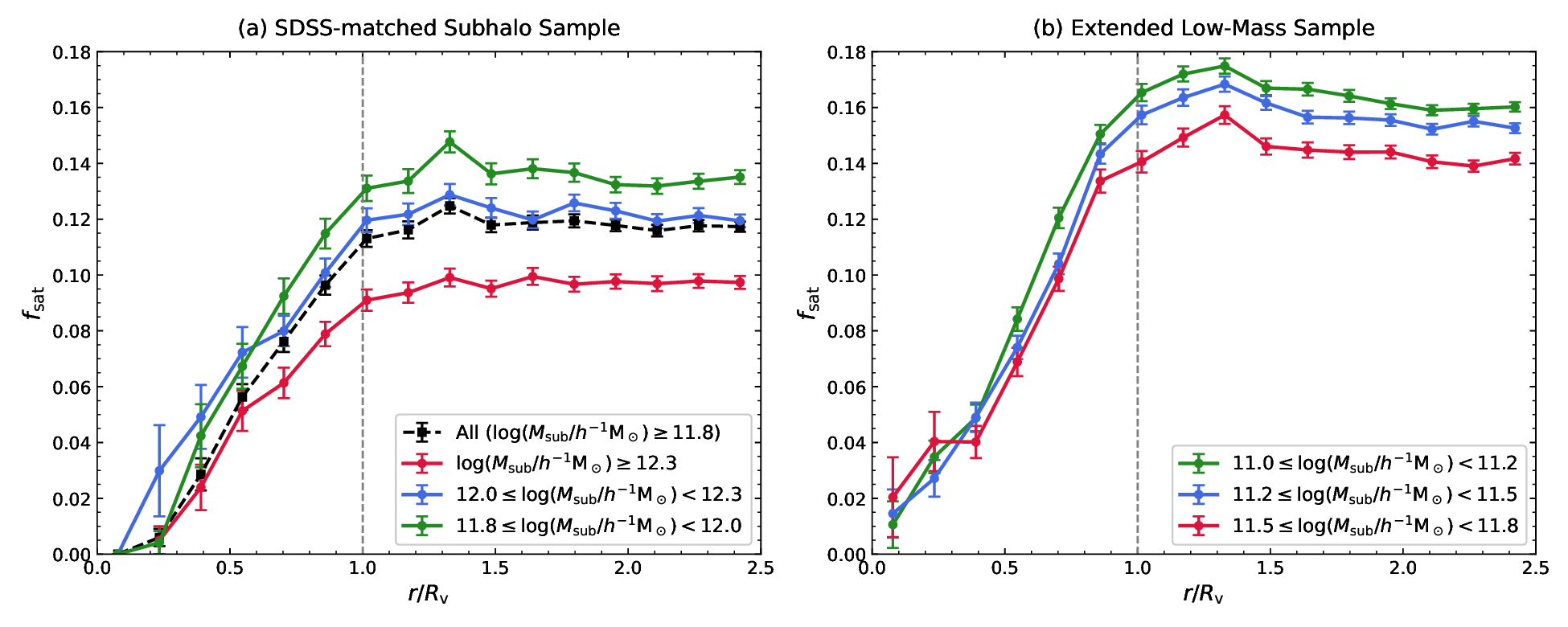}
    \caption{Radial dependence of the satellite fraction, $f_{\rm sat}$,
    in ELUCID voids. Panel~(a) shows the SDSS-matched subhalo sample,
    and panel~(b) shows the extended low-mass sample. The satellite
    fraction is defined as $N_{\rm sat}/(N_{\rm sat}+N_{\rm cen})$
    within each radial bin for each individual void, and then averaged
    over all contributing voids. Error bars denote the standard error
    of the mean across voids. The vertical dashed line marks the void
    boundary at $r/R_{\rm v}=1$.}
    \label{fig:void_satellite_fraction}
\end{figure}

To characterize the radial variation of the composition of the subhalo
population, we measure the satellite fraction,
$f_{\rm sat}=N_{\rm sat}/(N_{\rm sat}+N_{\rm cen})$, where central
subhaloes are the main subhaloes of their FOF haloes and the others are
satellites. Figure~\ref{fig:void_satellite_fraction} shows $f_{\rm sat}$
as a function of $r/R_{\rm v}$ for the SDSS-matched sample (panel~a) and
the extended low-mass sample (panel~b), computed void by void as
described in Section~\ref{subsec:statistics}.

The satellite fraction depends strongly on void-centric distance in both
samples. In the SDSS-matched sample, no satellites are found in the
innermost bin, and the fraction of the full sample increases to
approximately $0.03$ at $r/R_{\rm v}\simeq0.4$ and $0.11$ at the void
boundary, with a maximum of about $0.13$ at $r/R_{\rm v}\simeq1.3$, and
then decreases slightly to $\simeq0.12$ at larger radii. The extended
low-mass sample shows a similar pattern, with $f_{\rm sat}\lesssim0.04$
at $r/R_{\rm v}\lesssim0.4$, $0.14$--$0.17$ at the void boundary, a
maximum near $r/R_{\rm v}\simeq1.3$ of about $0.18$, $0.17$, and
$0.16$ for the three mass intervals from the lowest to the highest mass,
and a decrease by about $0.02$ to $\simeq0.16$, $0.15$, and $0.14$ at
larger radii.

The satellite fraction also depends on subhalo mass. At $r/R_{\rm v}\gtrsim0.7$,
the satellite fraction is higher for
lower-mass subhaloes. In the SDSS-matched sample, the maximum is about
$0.15$ for the lowest-mass interval,
$11.8\leq\log(M_{\rm sub}/h^{-1}M_\odot)<12.0$, compared with about
$0.10$ for the highest-mass interval,
$\log(M_{\rm sub}/h^{-1}M_\odot)\geq12.3$. The same ordering is seen in
the extended sample, with smaller differences. At smaller radii, the
ordering is not systematic, within large uncertainties. 

The decrease of the satellite fraction toward void interiors is
consistent with the environmental dependence of the satellite fraction
across the cosmic web, which decreases from nodes to filaments, walls,
and voids, particularly for low-mass galaxies \cite{Zhang2021b}. A
similar radial trend was found by the Euclid Collaboration, who reported
an increasing satellite fraction with increasing void-centric distance
for galaxies in a semi-analytic model applied to the Millennium
simulation in a Euclid-like lightcone \cite{Euclid2026}. While their
analysis concerns galaxies, and ours the satellite fraction of subhaloes
in the ELUCID simulation, both show a qualitatively similar radial
dependence. 

\subsubsection{Subhalo formation redshift}
\label{sec:subhalo_zf}

\begin{figure}[htbp]
    \centering
    \includegraphics[width=1.0\textwidth]{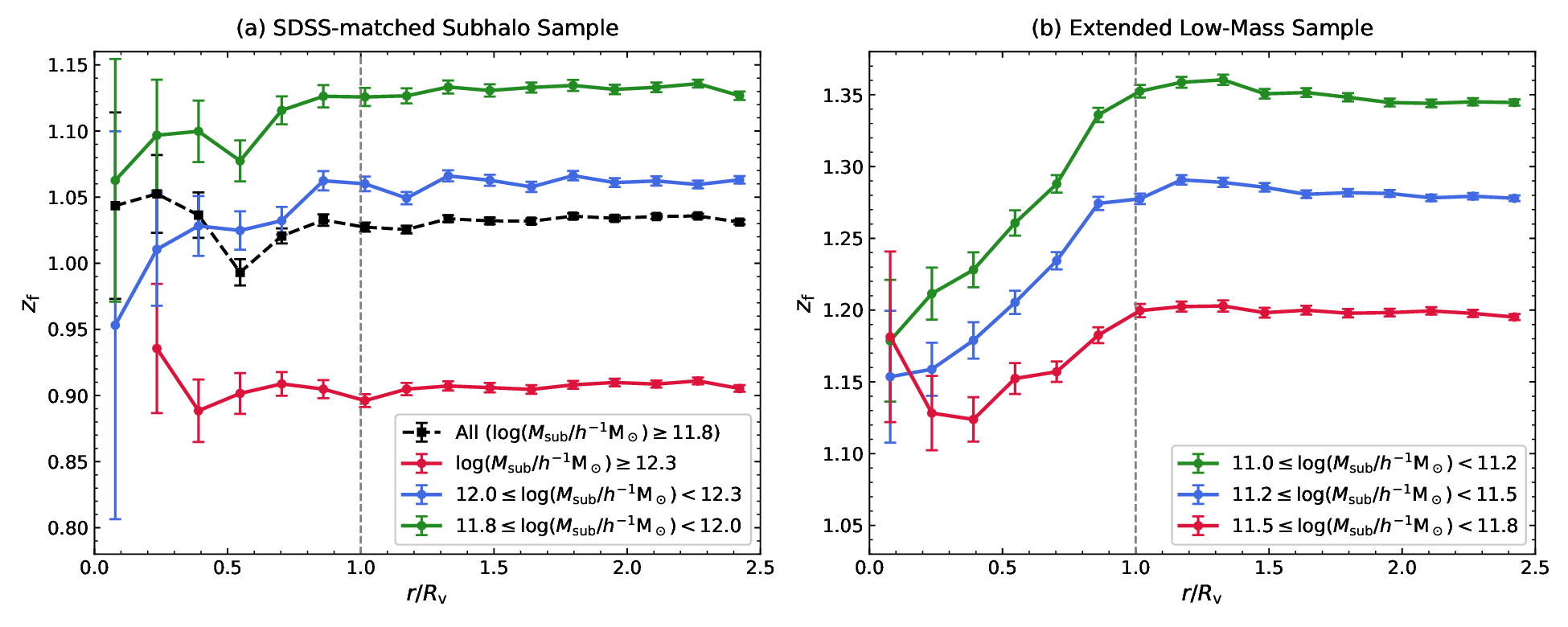}
    \caption{Radial dependence of the mean subhalo formation redshift,
    $z_{\rm f}$, in ELUCID voids. Panel~(a) shows the SDSS-matched
    subhalo sample, divided into three approximately equal-number mass
    subsamples above $\log(M_{\rm sub}/h^{-1}M_\odot)=11.8$, while
    panel~(b) extends the analysis to lower-mass subhaloes using three
    fixed mass intervals. Error bars denote the standard error of the
    mean across voids. The vertical dashed line marks the void boundary
    at $r/R_{\rm v}=1$. Only radial bins contributed by at least 10 voids
    are shown.}
    \label{fig:void_subhalo_zf}
\end{figure}

Figure~\ref{fig:void_subhalo_zf} shows the radial dependence of the
subhalo formation redshift, $z_{\rm f}$, for the mass subsamples defined
at the beginning of Section~\ref{sec:halo_properties}.

For the SDSS-matched sample, the mean formation redshift of the full
sample shows only a weak radial dependence, varying from approximately
$z_{\rm f}\simeq1.04$ in the inner void region to
$\simeq1.03$--$1.04$ outside the void boundary. The mass subsamples
reveal a strong dependence on subhalo mass: at $r/R_{\rm v}\gtrsim1$,
$z_{\rm f}\simeq0.90$, $1.06$, and $1.13$ for the high-, intermediate-,
and low-mass subsamples, respectively. Within the void interior, the
two lower-mass subsamples show an increase of $z_{\rm f}$ of about
$0.03$--$0.05$ from $r/R_{\rm v}\simeq0.2$ to the void boundary, whereas
the highest-mass subsample is nearly constant. The innermost bins
($r/R_{\rm v}\lesssim0.15$) have large uncertainties and are not used to
characterize the trends. All three subsamples become approximately
constant outside the void boundary. The nearly flat profile of the full
sample is consistent with the increase of $z_{\rm f}$ at fixed mass being
partly offset by the increase of the mean subhalo mass toward the void
boundary (Section~\ref{sec:subhalo_mass}), since more massive subhaloes
have lower formation redshifts.

We further examined subhaloes with
$\log(M_{\rm sub}/h^{-1}M_\odot)\geq12.3$ by dividing them into
five mass intervals of $0.3$ dex width (with the highest interval
extending to $\geq13.5$). At $r/R_{\rm v}\gtrsim0.4$ (and
$r/R_{\rm v}\gtrsim0.55$ for the highest-mass interval), the radial
variation in $z_{\rm f}$ is at most about $0.05$ in all five intervals,
consistent with no significant radial dependence. Meanwhile, $z_{\rm f}$
continues to decrease with increasing subhalo mass, from about $0.98$
to $0.63$. The innermost void regions contain few massive subhaloes,
so these intervals do not constrain the radial dependence at
$r/R_{\rm v}\lesssim0.4$.

Panel~(b) shows a more pronounced trend. For
$11.0\leq\log(M_{\rm sub}/h^{-1}M_\odot)<11.2$, the mean formation
redshift increases from about $1.21$ at $r/R_{\rm v}\simeq0.2$ to
$\simeq1.35$ around the void boundary. A similar but weaker increase is
seen for $11.2\leq\log(M_{\rm sub}/h^{-1}M_\odot)<11.5$, from about
$1.16$ to $\simeq1.28$. The highest-mass interval,
$11.5\leq\log(M_{\rm sub}/h^{-1}M_\odot)<11.8$, also increases, from
about $1.13$ at $r/R_{\rm v}\simeq0.2$--$0.4$ to $\simeq1.20$ at the
boundary, although by a smaller amount. In all three intervals the
formation redshift becomes nearly constant beyond
$r/R_{\rm v}\simeq1$.

These trends are not caused by a change of the mean subhalo mass within
the mass intervals. The mean mass of the whole extended sample changes
by only about $0.04$ dex across the voids
(Figure~\ref{fig:void_subhalo_mass}), and the change within each
narrower interval is of similar or smaller size. Since $z_{\rm f}$
decreases by about $0.3$ per dex of subhalo mass in panel~(b), a change
of $0.04$ dex in the mean mass would shift $z_{\rm f}$ by only
$0.3\times0.04\simeq0.01$, much smaller than the observed increases of
$0.07$--$0.14$. Moreover, the mean mass increases toward the void
boundary, which would lower $z_{\rm f}$, opposite to the observed trend.

The satellite fraction increases toward the void boundary
(Section~\ref{sec:satellite_fraction}), and satellite subhaloes have higher
formation redshifts than central subhaloes of the same mass, so that part
of the increase of $z_{\rm f}$ of all subhaloes reflects the changing
central--satellite composition. For central subhaloes alone, $z_{\rm f}$
still increases toward the void boundary in the extended low-mass sample,
by about $0.09$, $0.08$, and $0.04$ for the three intervals, compared with
$0.14$, $0.12$, and $0.07$ for all subhaloes. In the SDSS-matched sample,
the increase for central subhaloes is $\lesssim0.03$.

The dependence of the halo formation time on the environment is mass
dependent. In simulations, low-mass void haloes are the youngest and
node haloes the oldest \cite{Hellwing2021}, and haloes in lower-density
regions, such as walls and voids, form later on average in IllustrisTNG
\cite{Martizzi2020}. For ELUCID subhaloes, there is a characteristic
subhalo mass of $\sim10^{12}\,h^{-1}M_\odot$ below which subhaloes in
nodes are older than subhaloes of the same mass in filaments, walls, or
voids, while above it the trend reverses \cite{Zhang2021b}. A similar
reversal is found for the stellar-to-halo mass ratio of GAMA central
galaxies, which was interpreted as low-mass haloes in nodes being older
than haloes of the same mass in voids, and high-mass haloes in nodes
being younger \cite{Tojeiro2017}. This is related to the assembly bias
of dark-matter haloes \cite{GaoLiang2005,GaoLiang2007}. In the
MDPL2-SAG catalogue, low-mass haloes in voids are younger than the
average halo population while massive void haloes are older, and
low-mass haloes occupy the inner regions of voids while more massive
haloes are located closer to the void walls \cite{Alfaro2025}. 

Our results reveal a mass-dependent radial trend in the formation
redshift. For lower-mass subhaloes, the formation redshift generally
increases with void-centric distance inside voids, with a stronger
trend at lower masses. For subhaloes with $M_{\rm sub}\gtrsim10^{12.3}\,h^{-1}M_\odot$,
we find no significant radial dependence over the radial range
adequately sampled by our data.

\subsubsection{Halo size}
\label{sec:halo_size}

\begin{figure}[htbp]
    \centering
    \includegraphics[width=1.0\textwidth]{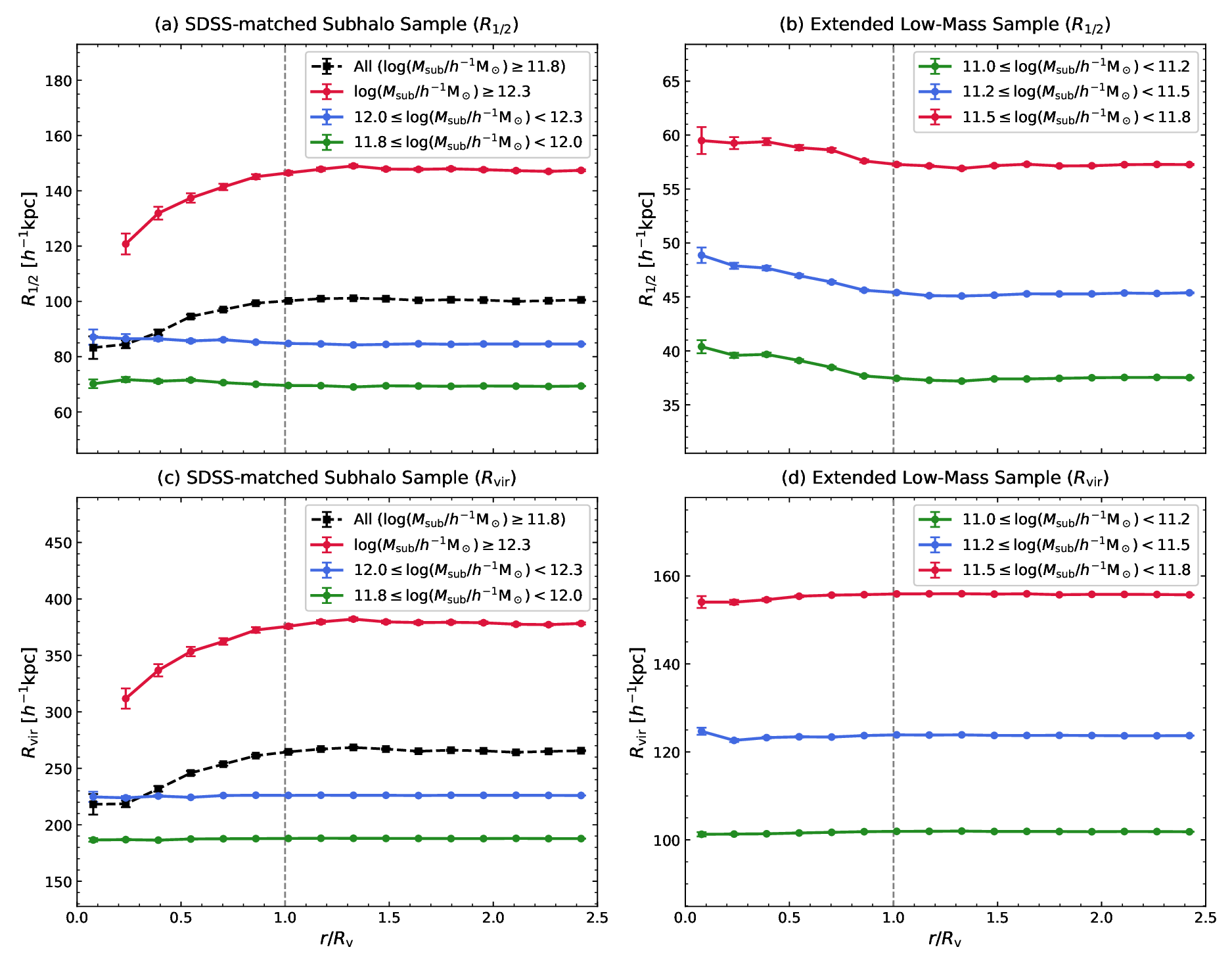}
    \caption{Radial dependence of halo size in ELUCID voids,
    characterized by the subhalo half-mass radius $R_{1/2}$ (top panels)
    and the virial radius $R_{\rm vir}$ of the host halo associated with
    central subhaloes (bottom panels). Panels~(a) and (c) show the
    SDSS-matched subhalo sample, while panels~(b) and (d) show the
    extended low-mass sample. Error bars denote the standard error of
    the mean across voids. The vertical dashed line marks the void
    boundary at $r/R_{\rm v}=1$.}
    \label{fig:void_halo_size}
\end{figure}

We next examine the radial dependence of halo size using two
characteristic radii: the subhalo half-mass radius, $R_{1/2}$, which
describes the internal mass distribution of an individual subhalo and is
measured for both central and satellite subhaloes, and the virial radius,
$R_{\rm vir}$, which is defined for the host halo associated with a
central subhalo and characterizes its overall extent. Since $R_{\rm vir}$
is defined through a fixed mean enclosed overdensity,
$M_{\rm vir}\propto R_{\rm vir}^3$, and $M_{\rm sub}$ is a useful proxy
for $M_{\rm vir}$ for central subhaloes (Section~\ref{sec:elucid}),
any radial trend in $R_{\rm vir}$ within a fixed $M_{\rm sub}$ interval
is expected to be driven primarily by variations in the mean subhalo
mass within that interval.

For the half-mass radius of the SDSS-matched sample in panel~(a), the mean
$R_{1/2}$ of the full sample increases from about
$84\,h^{-1}\,\mathrm{kpc}$ near the void centre to
$100\,h^{-1}\,\mathrm{kpc}$ around the void boundary, and remains nearly
constant outside the void. This apparent increase is mainly driven by the
highest-mass subsample: for $\log(M_{\rm sub}/h^{-1}M_\odot)\geq12.3$, the
mean $R_{1/2}$ increases from approximately $121$ at
$r/R_{\rm v}\simeq0.2$ to $146\,h^{-1}\,\mathrm{kpc}$ at the void
boundary. In contrast, the two lower-mass subsamples show little
variation, with a weak decrease of about $2$--$3\%$ from
$r/R_{\rm v}\simeq0.2$ to the boundary: the
$12.0\leq\log(M_{\rm sub}/h^{-1}M_\odot)<12.3$ subsample goes from about
$86.5$ to $84.8\,h^{-1}\,\mathrm{kpc}$, and the
$11.8\leq\log(M_{\rm sub}/h^{-1}M_\odot)<12.0$ subsample from about
$71.7$ to $69.6\,h^{-1}\,\mathrm{kpc}$.

The extended low-mass sample in panel~(b) provides a complementary view.
$R_{1/2}$ decreases toward the void boundary in all three intervals. From
$r/R_{\rm v}\simeq0.2$ to the boundary, it goes from about $39.6$ to
$37.5\,h^{-1}\,\mathrm{kpc}$ for
$11.0\leq\log(M_{\rm sub}/h^{-1}M_\odot)<11.2$ ($\sim5\%$), from about
$47.9$ to $45.4\,h^{-1}\,\mathrm{kpc}$ for
$11.2\leq\log(M_{\rm sub}/h^{-1}M_\odot)<11.5$ ($\sim5\%$), and from about
$59.3$ to $57.3\,h^{-1}\,\mathrm{kpc}$ for
$11.5\leq\log(M_{\rm sub}/h^{-1}M_\odot)<11.8$ ($\sim3\%$), becoming
approximately constant beyond the void boundary. Subhaloes of a given
mass are therefore slightly larger in the inner regions of the voids, by
about $2$--$5\%$ over $11.0\leq\log(M_{\rm sub}/h^{-1}M_\odot)<12.3$.
This weak trend is largely a consequence of the mixture of central and
satellite subhaloes. The mean $R_{1/2}$ of all subhaloes in an interval
equals that of the central subhaloes within about $1\%$ at
$r/R_{\rm v}\simeq0.2$, but is lower by $3$--$5\%$ at
$r/R_{\rm v}\gtrsim1$, which reflects the increasing fraction of
satellite subhaloes (Section~\ref{sec:satellite_fraction}), with smaller half-mass radii, toward the void
boundary. For central subhaloes alone, $R_{1/2}$ varies by about $1\%$ or less
from $r/R_{\rm v}\simeq0.2$ to the boundary in the five intervals other
than the open-ended $\log(M_{\rm sub}/h^{-1}M_\odot)\geq12.3$ one (for
example from $71.7$ to $72.4\,h^{-1}\,\mathrm{kpc}$ for
$11.8\leq\log(M_{\rm sub}/h^{-1}M_\odot)<12.0$ and from $40.0$ to
$39.6\,h^{-1}\,\mathrm{kpc}$ for
$11.0\leq\log(M_{\rm sub}/h^{-1}M_\odot)<11.2$), with no systematic
sign. At fixed mass, the half-mass radius of central subhaloes therefore
shows no significant dependence on void-centric distance.

For the virial radius of the host haloes of central subhaloes, the
behaviour in panel~(c) is similar to that of $R_{1/2}$ in panel~(a). The
mean $R_{\rm vir}$ of the full sample increases from about
$218\,h^{-1}\,\mathrm{kpc}$ in the inner void region to
$265\,h^{-1}\,\mathrm{kpc}$ near the void boundary, followed by an
approximately constant value outside. The radial variation is again
concentrated in the highest-mass subsample, for which $R_{\rm vir}$
increases from about $312$ at $r/R_{\rm v}\simeq0.2$ to
$376\,h^{-1}\,\mathrm{kpc}$ at the void boundary, while the two
lower-mass subsamples remain close to $226$ and
$188\,h^{-1}\,\mathrm{kpc}$, with variations below $1\%$. In
panel~(d), $R_{\rm vir}$ is nearly constant across the voids, increasing
by only about $1\%$ from $r/R_{\rm v}\simeq0.2$ to the boundary
($101.3\to101.9$, $122.6\to123.9$, and
$154.0\to155.9\,h^{-1}\,\mathrm{kpc}$ for the three intervals), which is
consistent with the small drift of the mean mass. The lowest-,
intermediate-, and highest-mass intervals stay close to $102$, $124$,
and $156\,h^{-1}\,\mathrm{kpc}$.

The apparent increase of $R_{1/2}$ and $R_{\rm vir}$ for the
highest-mass subsample in panels~(a) and (c) should be interpreted with
caution. The broad $\log(M_{\rm sub}/h^{-1}M_\odot)\geq12.3$ interval
spans more than two orders of magnitude in subhalo mass, and its mean
mass increases toward the void boundary. Given the strong dependence of
halo size on mass, this internal mass variation can naturally produce an
apparent radial trend, as for the high-mass elliptical galaxies in
Section~\ref{sec:galaxy_size}. To test this, we divide the high-mass
sample into narrower mass intervals. The apparent increase is largely
absent within these intervals: both $R_{1/2}$ and $R_{\rm vir}$ remain
approximately constant with void-centric distance over the intervals that
contain sufficient tracers across the void interior, while the
highest-mass intervals become sparse toward small $r/R_{\rm v}$ and do
not constrain the radial dependence inside the void.

Overall, the apparent radial increase in halo size for subhaloes with
$\log(M_{\rm sub}/h^{-1}M_\odot)\geq12.3$ is primarily caused by mass
mixing within this broad mass interval. At fixed subhalo mass, both
$R_{1/2}$ and $R_{\rm vir}$ show little or no significant dependence on
void-centric distance over the adequately sampled radial range. The weak
decrease of $R_{1/2}$ toward the void boundary in the extended low-mass
sample, and in the two lower-mass intervals of the SDSS-matched sample, is
largely removed when only central subhaloes are considered, and is
therefore largely accounted for by the changing central--satellite
composition.

\subsubsection{Halo concentration}
\label{sec:halo_concentration}

\begin{figure}[htbp]
\centering
\includegraphics[width=1.0\textwidth]{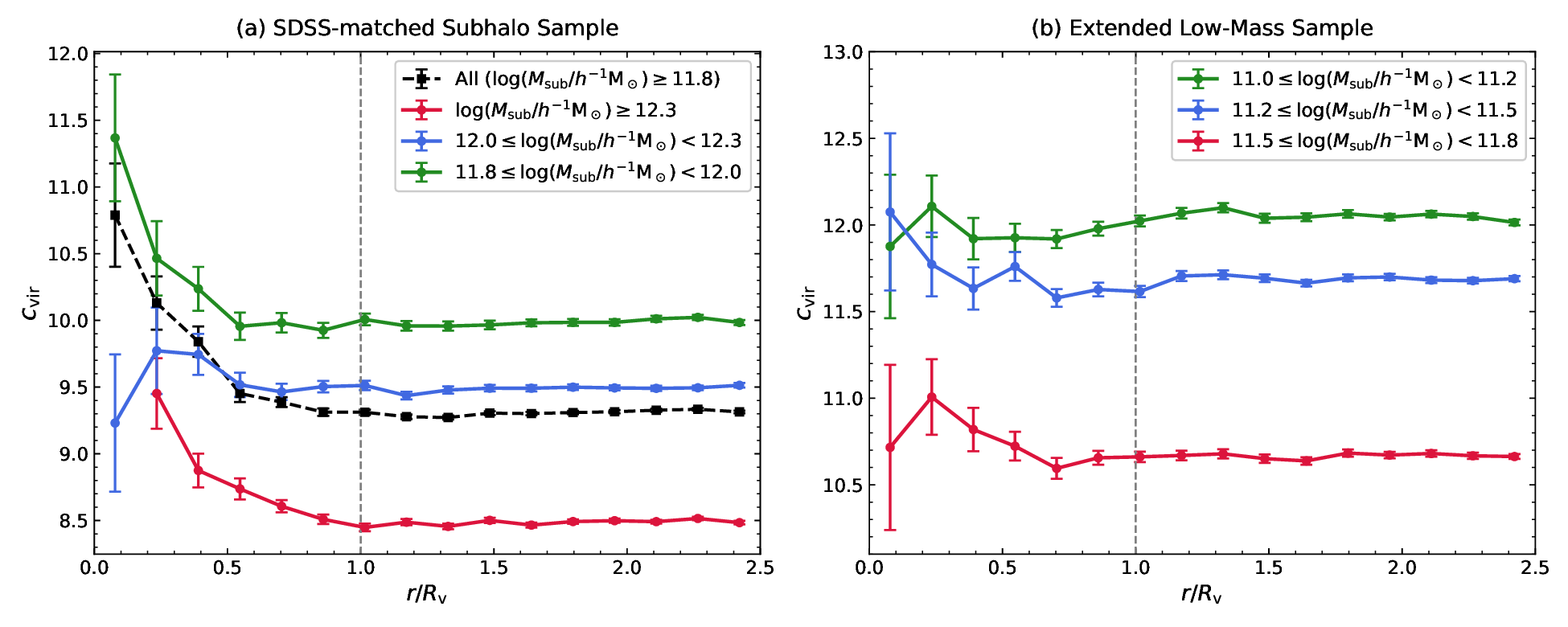}
\caption{Mean virial concentration $c_{\rm vir}$ of the host haloes associated
with central subhaloes as a function of void-centric distance. Panel~(a)
shows the SDSS-matched subhalo sample with
$\log(M_{\rm sub}/h^{-1}M_\odot)\geq 11.8$, divided into three mass
intervals, while panel~(b) shows the extended low-mass sample. Error
bars denote the standard error of the mean across voids. The vertical
dashed line marks the void boundary at $r/R_{\rm v}=1$.
}
\label{fig:halo_cvir}
\end{figure}

We next examine the virial concentration of the host haloes associated
with central subhaloes. As defined in Section~\ref{sec:elucid},
$c_{\rm vir}=R_{\rm vir}/r_{\rm s}$, where $r_{\rm s}$ is the NFW scale
radius. Figure~\ref{fig:halo_cvir} shows the mean $c_{\rm vir}$ as a
function of void-centric distance for the SDSS-matched sample in
panel~(a) and the extended low-mass sample in panel~(b).

The concentration depends strongly on subhalo mass. Outside the voids,
the SDSS-matched intervals have $c_{\rm vir}\simeq9.98$, $9.49$, and
$8.49$ for $11.8\leq\log(M_{\rm sub}/h^{-1}M_\odot)<12.0$,
$12.0\leq\log(M_{\rm sub}/h^{-1}M_\odot)<12.3$, and
$\log(M_{\rm sub}/h^{-1}M_\odot)\geq12.3$, respectively, and the three
intervals of the extended sample have $c_{\rm vir}\simeq12.05$, $11.69$,
and $10.67$ with increasing mass. Lower-mass subhaloes are therefore
associated with more concentrated host haloes over the whole mass range.

Within the voids, $c_{\rm vir}$ also varies with void-centric distance.
In panel~(a), the highest-mass interval decreases from about $9.45$ at
$r/R_{\rm v}\simeq0.2$ to $\simeq8.45$ at the void boundary. The
$11.8\leq\log(M_{\rm sub}/h^{-1}M_\odot)<12.0$ interval decreases from
about $10.5$ at $r/R_{\rm v}\simeq0.2$ to $\simeq9.96$ at
$r/R_{\rm v}\simeq0.55$, and the
$12.0\leq\log(M_{\rm sub}/h^{-1}M_\odot)<12.3$ interval shows a weaker
decrease from about $9.77$ at $r/R_{\rm v}\simeq0.2$ to $\simeq9.46$ at
$r/R_{\rm v}\simeq0.7$. In all three intervals, $c_{\rm vir}$ is
approximately constant outside the void boundary. The full sample
decreases from about $10.8$ to $9.3$, partly because the mean subhalo
mass increases toward the void boundary
(Section~\ref{sec:subhalo_mass}) while $c_{\rm vir}$ decreases with mass.

The decrease for the broad $\log(M_{\rm sub}/h^{-1}M_\odot)\geq12.3$
interval may be partly due to the same mass mixing as discussed for the
halo size (Section~\ref{sec:halo_size}), since $c_{\rm vir}$ decreases
with mass by about $2$ per dex between the lower intervals. The narrower
intervals are much less affected: a change of $0.03$ dex in the mean mass
within an interval of $0.2$--$0.3$ dex width would change $c_{\rm vir}$
by only $\simeq0.06$, much smaller than the observed variations.

In the extended low-mass sample in panel~(b), the radial variation is
weaker and less regular. For $11.5\leq\log(M_{\rm sub}/h^{-1}M_\odot)<11.8$,
$c_{\rm vir}$ decreases from about $11.0$ at $r/R_{\rm v}\simeq0.2$ to
$\simeq10.6$ at $r/R_{\rm v}\simeq0.7$ and is $\simeq10.67$ beyond the
void boundary. The $11.2\leq\log(M_{\rm sub}/h^{-1}M_\odot)<11.5$
interval fluctuates between $\simeq11.6$ and $11.8$ at
$r/R_{\rm v}\lesssim0.6$, has its lowest values of $\simeq11.6$ at
$r/R_{\rm v}\simeq0.7$--$1$, and returns to $\simeq11.7$ outside the
void. For $11.0\leq\log(M_{\rm sub}/h^{-1}M_\odot)<11.2$, $c_{\rm vir}$
is, if anything, slightly lower in the void interior ($\simeq11.92$ at
$r/R_{\rm v}\simeq0.4$--$0.7$ compared with $\simeq12.05$ outside). The
amplitudes are at most about $3\%$ and comparable to the uncertainties.
Subhaloes near the lower mass limit contain only a few hundred particles,
so their structural parameters may be affected by sampling and fitting
uncertainties (Section~\ref{subsec:subhalo_samples}), and we treat the
variation in this mass range with caution.

Overall, the subhaloes with $\log(M_{\rm sub}/h^{-1}M_\odot)\geq11.8$
are more concentrated toward the void interior, whereas the radial
dependence is weak and irregular in the extended low-mass sample. This
behaviour differs from that of the formation redshift
(Section~\ref{sec:subhalo_zf}). The radial dependence of $z_{\rm f}$ is
strong in the extended low-mass sample and weakens with increasing
mass, becoming insignificant for
$M_{\rm sub}\gtrsim10^{12.3}\,h^{-1}M_\odot$. The mass dependences of
the radial trends in $c_{\rm vir}$ and $z_{\rm f}$ therefore differ, and
the origin of this difference is not explored further in this work.

Our previous study of haloes around cosmic filaments found that the radial 
dependence of central-subhalo concentration varies among filament catalogues
constructed using different smoothing scales and thresholds \cite{Zhang2025}.
For the most prominent filament catalogue, the concentration of central 
subhaloes with $M_{\rm sub}>10^{11}\,h^{-1}M_\odot$ increases toward the 
filament spines, whereas this trend is weaker or absent for other catalogues. 
Using a cosmic-web classification into voids, walls, filaments, and nodes, 
\cite{Hellwing2021} found that environmental differences in halo 
concentration are most pronounced below a characteristic mass 
of $\simeq6\times10^{10}\,h^{-1}M_\odot$, with void haloes having 
the lowest concentrations, followed by haloes in walls, filaments, and
nodes. These studies examine concentration variations around filaments
or across distinct cosmic-web environments, whereas our analysis 
focuses on radial profiles within voids. The results therefore probe
different aspects of the environmental dependence of halo concentration.

\subsubsection{Halo spin}
\label{sec:halo_spin}

\begin{figure}[htbp]
\centering
\includegraphics[width=1.0\textwidth]{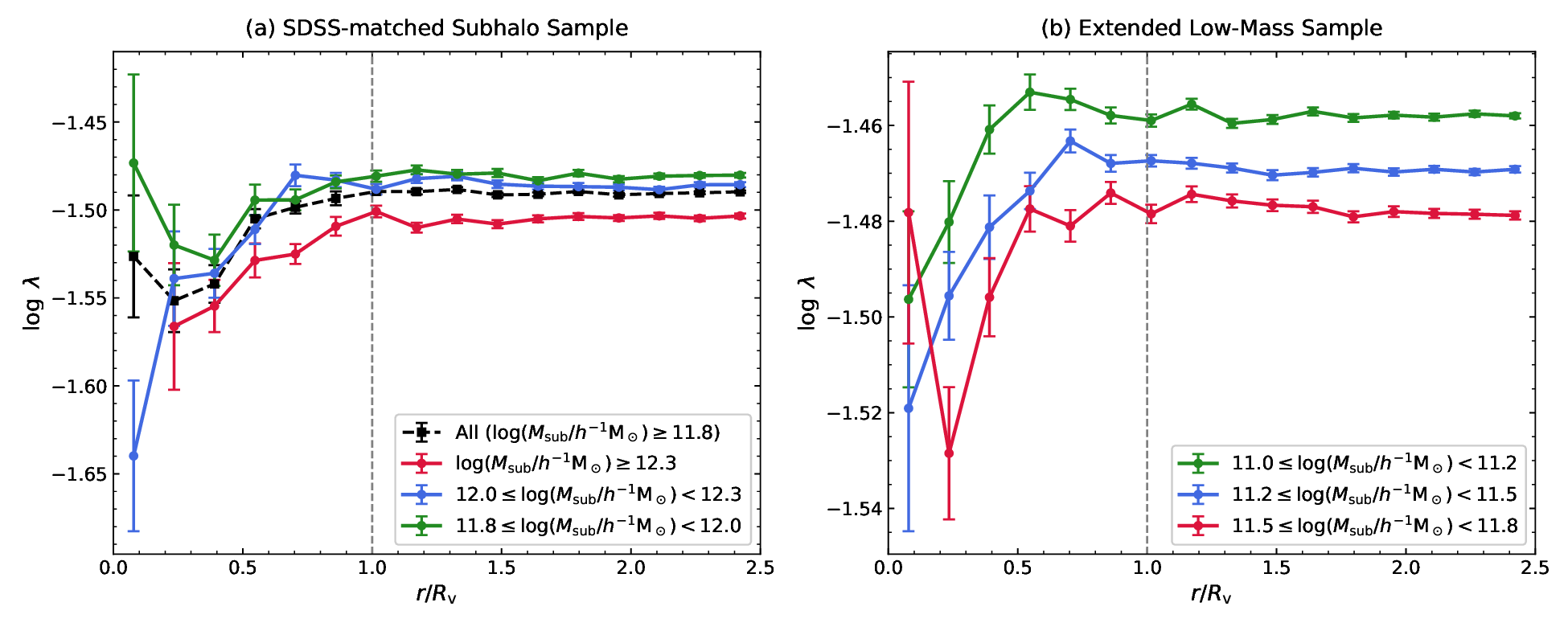}
\caption{
Mean subhalo spin parameter $\log \lambda$ as a function of void-centric
distance. Panel~(a) shows the SDSS-matched subhalo sample with
$\log(M_{\rm sub}/h^{-1}M_\odot)\geq 11.8$, divided into three mass
intervals, while panel~(b) shows the extended low-mass sample. Error
bars denote the standard error of the mean across voids. The vertical
dashed line marks the void boundary at $r/R_{\rm v}=1$.}
\label{fig:subhalo_spin}
\end{figure}

We next examine the dimensionless spin parameter $\lambda$ of subhaloes,
defined in Section~\ref{sec:elucid}. Since $\lambda$ is calculated from
the mass, angular momentum, and energy of the subhalo, it provides a
measure of the rotational support of its dark-matter structure.
Figure~\ref{fig:subhalo_spin} shows the mean $\log\lambda$ as a function
of void-centric distance for the SDSS-matched sample (panel~a) and the
extended low-mass sample (panel~b).

The spin depends weakly on subhalo mass. At $r/R_{\rm v}\gtrsim1$,
lower-mass subhaloes have slightly higher spin parameters: in the
SDSS-matched sample, $\log\lambda\simeq-1.480$, $-1.485$, and $-1.505$ for
the lowest-, intermediate-, and highest-mass intervals, and in the
extended sample $\log\lambda\simeq-1.458$, $-1.469$, and $-1.477$ with
increasing mass. The mean masses of adjacent intervals differ by
$\Delta\langle\log M_{\rm sub}\rangle\simeq0.24$ and $0.29$ dex in the
extended sample and by $0.24$ and $0.60$ dex in the SDSS-matched sample,
while their spins differ by only $\Delta\log\lambda\simeq0.005$--$0.02$.
The dependence of the spin on mass is therefore weak, with a slope
$\Delta\log\lambda/\Delta\langle\log M_{\rm sub}\rangle$ of about $-0.03$
per dex, and the spin changes by only $\Delta\log\lambda\simeq0.02$--$0.03$ 
over the whole mass range of each sample.

The radial dependence is larger than this mass dependence. In both
panels, at fixed subhalo mass, $\log\lambda$ is lower in the void
interiors than near and outside the void boundary. In panel~(a), it
increases from $r/R_{\rm v}\simeq0.2$ to the void boundary by
$\Delta\log\lambda\simeq0.04$, $0.05$, and $0.07$ for the lowest-,
intermediate-, and highest-mass intervals, from approximately $-1.52$,
$-1.54$, and $-1.57$ to $-1.48$, $-1.49$, and $-1.50$. In panel~(b), the
increase is $\Delta\log\lambda\simeq0.02$, $0.03$, and $0.05$ for the
three intervals with increasing mass, from approximately $-1.48$, $-1.50$,
and $-1.53$ to $-1.46$, $-1.47$, and $-1.48$. Most of the change occurs at
$r/R_{\rm v}\lesssim0.7$, and the spin is approximately constant at
$r/R_{\rm v}\gtrsim1$, apart from a shallow maximum of about $0.005$ dex at
$r/R_{\rm v}\simeq0.5$--$0.7$ in some intervals. The innermost bins
($r/R_{\rm v}\lesssim0.15$) contain only $12$--$126$ voids, have large
uncertainties, and are not monotonic. The change of the spin from the void
interior to the boundary at fixed mass, $\Delta\log\lambda\simeq0.02$--$0.07$,
is thus comparable to or larger than the change of
$\Delta\log\lambda\simeq0.02$--$0.03$ over the whole mass range of each
sample.

This radial dependence is not caused by the mass composition. The mean
mass of the SDSS-matched sample increases from $r/R_{\rm v}\simeq0.2$ to
the void boundary by $\Delta\langle\log M_{\rm sub}\rangle\simeq0.18$ dex. 
With the slope above, this would change
$\log\lambda$ by $0.18\times(-0.03)\simeq-0.005$, that is, lower the spin
toward the void boundary, opposite to and an order of magnitude smaller
than the observed increase of about $0.06$ of the full sample. The mean
mass within the two lower-mass intervals changes by only $0.006$ and
$0.011$ dex, and by $0.001$--$0.007$ dex within the intervals of the
extended sample, and by $0.18$ dex within the open-ended highest-mass
interval, so that mass mixing within the intervals changes $\log\lambda$
by at most about $0.005$, again much smaller than the observed increase.

The trend is also present for central subhaloes alone, and is slightly
stronger than for all subhaloes. From $r/R_{\rm v}\simeq0.2$ to the void
boundary, $\log\lambda$ of the central subhaloes increases by
$\Delta\log\lambda\simeq0.05$, $0.05$, and $0.07$ for the lowest-,
intermediate-, and highest-mass intervals of the SDSS-matched sample, and
by $0.04$, $0.04$, and $0.06$ for the three intervals of the extended
sample, compared with $0.04$, $0.05$, and $0.07$ and $0.02$, $0.03$, and
$0.05$ for all subhaloes. Satellite subhaloes have lower spins than
central subhaloes of the same mass, by about $0.02$--$0.10$ dex, and show no
systematic radial dependence at $r/R_{\rm v}\gtrsim0.7$. Since the satellite
fraction increases toward the void boundary
(Section~\ref{sec:satellite_fraction}), the changing central--satellite
composition lowers the spin of all subhaloes toward the boundary by
$0.003$--$0.008$ in the SDSS-matched intervals and by $0.010$--$0.016$ in
the extended intervals, which reduces, rather than produces, the radial
increase.

In simulations, the halo spin correlates with the cosmic-web environment,
being largest in filaments \cite{Ganeshaiah2021}, while the environmental
effect on the spin is much weaker than on the concentration and the
formation time, with more massive void and wall haloes having on average
lower spin \cite{Hellwing2021}. In observational data, the halo spins
estimated for HI-bearing galaxies are higher closer to filaments
\cite{Liu2026}. Our results are qualitatively consistent with these
findings and with our previous analysis of the spin around cosmic
filaments, where the spin decreases with increasing distance from the
filament axes \cite{Zhang2025}.

We emphasize that the present measurement is based directly on the
void-centric distance and does not attempt to separate the contributions
of the surrounding cosmic-web structures. The result therefore provides
an empirical characterization of the spin dependence on the void
environment rather than a determination of its underlying physical
mechanism.

\section{Conclusions}
\label{sec_conclusions}

We have investigated the dependence of galaxy and dark-matter subhalo
properties on void-centric distance by combining the SDSS DR7
volume-limited galaxy sample with the ELUCID constrained simulation.
Voids were identified independently in the galaxies and subhaloes using
the same \texttt{REVOLVER} procedure, yielding 389 SDSS voids and 404
ELUCID voids. Each property was measured on a void-by-void basis, for
both the full samples and narrower mass ranges. Our main findings are
summarized below.

For the SDSS galaxies:

\begin{enumerate}

\item \textbf{Stellar mass shows clear mass segregation.}
The mean stellar mass increases from the void interiors to the
boundaries and remains approximately constant outside the voids.

\item \textbf{Colour and sSFR retain weak radial trends at fixed stellar
mass.}
In the full sample, galaxies become slightly redder and their sSFR
decreases toward the void boundaries. Both trends are substantially
weaker within the stellar-mass subsamples, although a residual tendency
toward bluer colours and higher sSFR in the void interiors remains.

\item \textbf{Morphology shows little radial dependence at fixed stellar
mass.}
The spiral fraction slightly decreases in the full sample but remains
nearly constant within the stellar-mass subsamples. The overall decline
can largely be reproduced by the changing stellar-mass composition.

\item \textbf{The radial dependence of galaxy size is weak at fixed
stellar mass.}
The mean size of elliptical galaxies increases toward the void
boundaries, but this trend largely disappears for high-mass ellipticals
when narrower stellar-mass bins are used. Spiral galaxies show only weak
radial variations.

\end{enumerate}

For the ELUCID subhaloes:

\begin{enumerate}

\item \textbf{Subhalo mass and satellite fraction increase toward the
void boundaries.}
The mean subhalo mass increases from the void interiors to the
boundaries, similar to the radial dependence of the galaxy stellar mass.
The satellite fraction increases from nearly zero in the void interiors
to a roughly constant level near and beyond the void boundaries.

\item \textbf{The radial dependence of the formation redshift is
mass-dependent.}
Within fixed mass intervals, the mean formation redshift $z_{\rm f}$
increases toward the void boundaries for lower-mass subhaloes, that is,
subhaloes in the void interiors form later, and the change is larger at
lower masses. For subhaloes with
$M_{\rm sub}\gtrsim10^{12.3}\,h^{-1}M_\odot$, $z_{\rm f}$ shows no
significant radial dependence over the adequately sampled radial range.
The increase also persists for central subhaloes
alone, although it is smaller than for all subhaloes, because the
changing central--satellite composition contributes part of it.

\item \textbf{Halo concentration is higher in the void interiors for
$M_{\rm sub}\geq10^{11.8}\,h^{-1}M_\odot$.}
The host haloes associated with central subhaloes are more concentrated
in the void interiors within the narrower mass intervals of the
SDSS-matched sample. The radial dependence is weaker and less regular in
the extended low-mass sample.

\item \textbf{Halo size shows little dependence on void-centric distance
at fixed mass.}
The apparent increase in size in the broad highest-mass interval is
largely due to mass mixing. The half-mass radius of central subhaloes
shows no significant radial dependence at fixed mass, while the weak
radial variation of the half-mass radius of the full subhalo population
is largely accounted for by the changing central--satellite composition.
The virial radius is also nearly constant within narrower mass intervals.

\item \textbf{Subhalo spin increases from the void interiors to the
boundaries at fixed mass.}
The spin depends only weakly on subhalo mass, and its radial variation
within fixed mass intervals is larger than this mass dependence. The
trend is also found for central subhaloes alone, and neither the mass
dependence nor the changing central--satellite composition can account
for the observed radial increase.

\end{enumerate}

Overall, these results show that the radial dependence of galaxy and
subhalo properties varies considerably among the quantities examined. For
galaxies, stellar-mass segregation accounts for much of the apparent
radial dependence of morphology and size, while colour and sSFR retain
weaker trends at fixed stellar mass. For subhaloes, the formation
redshift at lower masses, the concentration in the SDSS-matched mass
range, and the spin retain radial dependence within fixed mass intervals,
whereas halo size shows little significant dependence. Thus, mass
segregation and the changing central--satellite composition are
important for interpreting the radial trends, but they do not account for
all of the variations in subhalo properties.

The void-centric distance used here is a geometric coordinate. The
voids are identified from the SDSS galaxy sample and the SDSS-matched
subhalo sample, respectively, and the properties of these samples are
measured in their corresponding voids. The extended low-mass subhalo
properties are measured in voids identified from the SDSS-matched
subhalo sample. We do not separate
the contributions of surrounding cosmic-web
structures or compare our samples with a control sample in
average-density environments. Our results therefore provide an empirical
characterization of property variations within voids, while the physical
origin of the residual trends at fixed mass, including possible
connections with assembly history and tidal environment, remains to be
investigated.

\acknowledgments

This research is funded by the National Natural Science Foundation of China (Nos. 12273088, 12633001).

Claude, ChatGPT, Gemini, and DeepSeek were used to assist with manuscript preparation, 
scientific discussion, and Python programming. The author takes full responsibility 
for the content of the manuscript.

This work is also supported by the High-Performance Computing Resource in 
the Core Facility for Advanced Research Computing at Shanghai Astronomical Observatory.

This work used {\tt Astropy:} a community-developed core Python package
and an ecosystem of tools and resources for astronomy \citep{Astropy2013, 
Astropy2018, Astropy2022}. 

Funding for the Sloan Digital Sky Survey IV has been provided by the
Alfred P. Sloan Foundation, the U.S. Department of Energy Office of
Science, and the Participating Institutions. SDSS acknowledges support
and resources from the Centre for High-Performance Computing at the
University of Utah. The SDSS website is www.sdss.org.

SDSS is managed by the Astrophysical Research Consortium for the
Participating Institutions of the SDSS Collaboration including the
Brazilian Participation Group, the Carnegie Institution for Science,
Carnegie Mellon University, the Chilean Participation Group, the
French Participation Group, Harvard-Smithsonian Center for
Astrophysics, Instituto de Astrof{\'i}sica de Canarias, The Johns
Hopkins University, Kavli Institute for the Physics and Mathematics of
the Universe (IPMU)/University of Tokyo, Lawrence Berkeley National
Laboratory, Leibniz Institut f{\"u}r Astrophysik Potsdam (AIP),
Max-Planck-Institut f{\"u}r Astronomie (MPIA Heidelberg),
Max-Planck-Institut f{\"u}r Astrophysik (MPA Garching),
Max-Planck-Institut f{\"u}r Extraterrestrische Physik (MPE), National
Astronomical Observatories of China, New Mexico State University, New
York University, University of Notre Dame, Observat{\'o}rio Nacional/
MCTI, The Ohio State University, Pennsylvania State University,
Shanghai Astronomical Observatory, United Kingdom Participation Group,
Universidad Nacional Aut{\'o}noma de M{\'e}xico, University of
Arizona, University of Colorado Boulder, University of Oxford,
University of Portsmouth, University of Utah, University of Virginia,
University of Washington, University of Wisconsin, Vanderbilt
University, and Yale University.


\bibliographystyle{JHEP}
\bibliography{biblio.bib}


\end{document}